\documentclass[useAMS,usenatbib,referee]{mn2e}
\usepackage{graphicx}
\usepackage{natbib}
\usepackage{times}
\usepackage{float}
\usepackage{amsmath}
\usepackage{epsfig}
\usepackage{xcolor}
\definecolor{red}{cmyk}{0,0,0,1}
\definecolor{blue}{cmyk}{0,0,0,1}
\definecolor{white}{cmyk}{0,0,0,0}

\newcommand{\drv}[2]{{{\mathrm{d} #1}\over {\mathrm{d} #2}}}

\newcommand{\im}{\mathrm{Im}}
\newcommand{\bsf}[1]{\mbox{\boldmath$\mathsf{#1}$}}

\begin{document}

\title[  
 Tidal interaction in the misaligned case ]
{The tidal excitation of $r$ modes  in a 
\textcolor{red}
{solar type} \textcolor{red}{star orbited by a}
 giant planet \textcolor{red}{companion}
 and the effect on orbital evolution II: The effect of tides in the  misaligned case}

\author[J. C. B. Papaloizou and  G.J. Savonije ]
{ J. C. B. Papaloizou $^{1}$\thanks{E-mail:
		J.C.B.Papaloizou@cam.ac.uk (JCBP)}, 
{G.J. Savonije$^{2}$\thanks{E-mail: g.j.savonije@uva.nl (GJS)},}
	    \\
	$^{1}$ DAMTP, Centre for Mathematical Sciences, University of
           Cambridge, Wilberforce Road, Cambridge CB3 0WA 	 \\    
	$^{2}$Anton Pannekoek Institute of Astronomy, University of Amsterdam, Science Park 904,
	NL-1098 XH, Amsterdam }

\maketitle
\maketitle

\date{Accepted. Received; in original form}

\pagerange{\pageref{firstpage}--\pageref{lastpage}} \pubyear{2010}


\label{firstpage}

{\textcolor{blue}{\begin{abstract}
We extend the study of Papaloizou \& Savonije of the tidal interactions of close orbiting  giant planets 
with a central solar type star to 
the situation where the spin axis of the central star and  the orbital angular momentum are misaligned.
We determine the tidal response taking into account the possibility of the excitation of $r$ modes and 
the effect of tidal forcing due to potential perturbations  which have  zero frequency in a  non rotating  frame. 
Although there is near resonance with  $r$ modes with degree $l'=1$ and orders $m=\pm1, $ 
half widths turn out to be sufficiently narrow so that in practice 
dissipation rates are found to be similar  to those produced by non resonant potential perturbations. 
We use our results to determine the evolution of the misalignment for the full range of initial inclination angles 
taking account of the spin down of the central star due to magnetic braking. 
Overall we find the rate of tidal evolution to be  unimportant for a one Jupiter mass planet with orbital period $\sim3.7d$
over a main sequence lifetime. 
However, it becomes significant for higher mass planets and shorter orbital periods, approximately scaling as 
the square of the planet mass and the inverse fourth power of the orbital period.
\end{abstract}}


\section{Introduction}
 \textcolor{blue}{ We extend the  study of the tidal interaction of a solar mass primary with a Jupiter
 mass secondary in a close circular orbit carried out by  \citet {PS23} (hereafter PS) to 
 consider the case when the stellar spin and orbital angular momentum vectors are misaligned.
 As in PS it is assumed  that turbulent viscosity  \citep[see eg.][]{Zahn1977,Duguid2020} operates
 in the stellar convective envelope  while the  spin  angular momentum of the planet is neglected.}
 
 \textcolor{blue}{ A number of physical processes have been invoked  in order to produce the initial distribution
  of alignment angles of close orbiting giant planets. These include quiescent phenomena such as disc migration that would
  lead to close alignment \citep[eg][]{LP86}, and dynamical interactions with a range of strengths \citep[see eg][and references therein]{Siegel}
  that may produce significant misalgnment. 
Subsequent to formation, tidal interactions
 can result in  orbital  evolution
leading to synchronisation of the orbital and component spins.
\citep[see][for a review]{O2014} and so affect the observed distribution.
Hence it is important to understand the extent to which this may have occurred
in order to understand conditions just post formation.}


 \textcolor{blue}{ In PS  the 
  spin and orbital angular momenta were assumed to be aligned. In particular  
 the spectrum of $r$ modes  associated with spherical harmonics of degree
 $l'=3$ and $l'=5$ was studied in detail.
It was found that  tidal interaction is unlikely to lead to significant orbital evolution away from these $r$ mode resonances which were found to be very narrow.
However, it may become significant if resonance can be maintained during tidal evolution. This may be possible as a result of the central star being spun  down
 through a process such as magnetic breaking. This counters the tendency of tidal interaction to spin up the star enabling the resonance to be maintained.
 Systems where this may have operated are Kepler 1643 and COROT 4 ( see discussion in PS and references therein).}
 

\textcolor{blue} {In this paper  we extend the calculations of PS to consider tidal interactions for which  the spin  and orbital angular momenta are misaligned.
  The aim is to use the results to   consider the effects of tidal interactions on  the spin-orbit alignment of close orbiting giant planets.
  This involves  determining the response to tidal forcing  associated with spherical harmonics of degree $l'=2$ and azimuthal mode numbers
  $m=\pm1,$ as well as for  $m=\pm 2$ which occurs in the aligned case. } 
  
  { When there is misalignment, and these forms of perturbing tidal potential are viewed in a frame that is aligned with and corotating with the central  star, }
  forcing frequencies $\Omega_s$ for $|m|=1,$ and $2\Omega_s$  for $|m|=2$ occur, with $2\pi/\Omega_s$ being its rotation period  respectively.
  The associated disturbances are stationary in a non rotating frame and may be associated with a strong inertial mode response in the convective envelope
  \citep[eg][]{PP81, OL2007, IP2010,  O2014} as well as an  $r$ mode response  in the radiative core (PS). These features mean that the  tidal dissipation in the misaligned case
  may not be simply related to that in the aligned case, an 
  aspect we are able to investigate.}
  
\textcolor{blue}{ As in PS we consider a simplified model for the rotating  primary star 
valid up to first order in $\Omega_s.$  In this approximation Coriolis forces  are retained but centrifugal forces neglected
such that the configuration is spherically symmetric. This means that centrifugal distortion
which would lead to the slow precession of the spin angular momentum vector around the total
angular momentum vector  is neglected.  The approximation can be viewed as having introduced a  background potential
to cancel out the centrifugal potential which remains fixed when the model is perturbed.
This situation results in the model  not possessing  the rigid tilt mode that exists without the approximation. 
The rigid tilt mode has $l'=1,|m|=1,$  and  eigenfrequency  zero in an inertial frame,  and though related to spin precession is not associated with tidal dissipation.
Notably both the approximate and full model are expected to have a spectrum of higher order $r$ modes
with eigenfrequencies, which when viewed in a frame rotating with the star,  have  magnitude  $\rightarrow \Omega_s $ with a relative correction that $\rightarrow 0$ as $\Omega_s \rightarrow 0$ \citep{PP78, Dewberry2023}.
Here we find that this spectrum does not play a dominant role on account of the  narrowness  of resonance widths (see Section \ref{Discuss}).
We use our results to investigate the effect of the tidal interaction in combination with magnetic breaking of the central star
on the evolution of the spin-orbit alignment over the main sequence life time.}

The plan of this paper is as follows.
In Section \ref{BasicM} we describe the basic configuration
adopted for the  system consisting of a primary star and orbiting planet.  The
coordinate systems defining the orbit and stellar frames together with notation used is given in \ref{Geometry}.
The development of the  perturbing tidal potential  acting on the primary in terms of spherical harmonics defined
in the two coordinate systems  with the aid of Wigner matrices is described in  Sections   \ref{pertpoten}-\ref{tidalpot}

We go on to consider the
 perturbation to the  external gravitational potential due to the primary and use it to determine the
 tidal forces acting on the orbit,  and  equations governing its evolution,  by making use of the Gauss equations  in Sections \ref{pertpot}-\ref{Gaussinc}.
Discussion of the  equations governing the evolution of the  semi-major axis and the angle between the spin and orbital angular momenta,
$\beta$, together with estimates of evolutionary time scales  are provided in Sections \ref{DiscGausseq}-\ref{testimate}.

The Numerical calculation of the tidal response of the star and the viscous dissipation in its convective envelope is outlined in Section \ref{Numerics},
with numerical results presented in Section  \ref{Numres}.
 These include  calculations of the fundamental $r$ mode resonance            
  and the general tidal response for a range of primary  rotation rates in Sections \ref{rmoderes} -\ref{Genresp}.

Numerical calculations of the evolution of the  spin orbit  alignment of the system incorporating the effects of magnetic braking  
 are considered in Sections \ref{Magb}-\ref{Numcalcevol}
 The effects  of increasing planet mass  and/or decreasing the orbital period, which speed up the tidal evolution rate,  are described  in   Section \ref{incMpef}.
 Finally we discuss our results and conclude in Section \ref{Discuss}.

\subsection{Basic configuration}\label{BasicM}

Following PS we consider a giant planet \textcolor{blue}{secondary of mass $M_p$  orbiting a star of mass $M_*$ taken to be} a solar mass described as
 the primary \footnote{As the secondary is treated as a point mass and the tidal response is linear, results can be scaled to apply to different masses (PS)}. 
Here, in an extension of the work of PS,  we consider  the situation where the orbital and spin angular momenta are misaligned.
Though, as in PS, we assume  efficient orbital circularisation potentially induced by tides assumed to act on the secondary,
thus limiting consideration to circular orbits.


\subsection{Coordinate system and notation}\label{Geometry}
Following  \cite{IP2021} (herafter IP) we define three Cartesian coordinate systems, each of them with origin $O$ at the centre of mass of the primary.
The $(X,'Y',Z')$ system, { or orbit frame,} is  such that  the orbital  angular momentum, ${\bf L}$, 
defines the direction of the $Z'$ axis. The  $X'$ and $Y'$ axes lie in the orthogonal orbital plane.  
The $(X,Y, Z)$ system, or stellar frame,  has $Z$ axis pointing in the direction of the stellar spin angular momentum ${\bf S}.$
We then choose the $Y$ axis to lie in the $(X',Y')$ plane with the angle between $Y$ and $Y'$
being  $\gamma.$ \footnote{This is defined in the sense of a positive rotation about $OZ'$ from $OY$  to $OY'.$}  
 The angle between ${\bf L}$ and ${\bf S}$ is $\beta.$ 
When the total angular momentum ${\bf J}={\bf L}+{\bf S}$ is conserved, it is natural to  consider
the $(X''Y''Z'')$ system, which we call the primary frame,  such that $Z''$ points in the direction of ${\bf J}.$
Naturally $OZ' ,OZ,$ and $OZ''$ lie in the same plane. The angle between $OZ'$ and $OZ''$ is $i.$

 The angle between the  line  of apsides associated with the assumed near Keplerian orbit  
 and the $X'$ axis is  denoted  by $\varpi.$
  Note that then  the angle between the apsidal line and the $Y$ axis,
  which can be chosen to  be the line of nodes  \footnote{ This is because $OZ'$ and $OZ''$ can be generated from $OZ$
by rotations about $OY$ and are thus normal to it.} is
given by  $ \varpi +\gamma - \pi/2$.
{ The coordinate systems together with the angles $\gamma$, $\beta$ and $\varpi$ are illustrated in Fig. 1 of IP.}  

\section{The perturbing tidal potential}\label{pertpoten}
The perturbing potential, $U$, can be expressed 
in spherical polar coordinates $(r,\theta', \phi')$ defined in the orbit frame $(X',Y',Z')$ with { origin at the centre of mass of primary(see IP and  PS).  
 As usual $r$ is the distance to the origin, $\theta'$ is the angle between the radius vector and the $Z'$ axis and $\phi'$ is the azimuthal angle.}
  We have
up to $O(a^{-3})$
\begin{equation}
 U=-\frac{GM_pr^2}{a^3}P_2(\cos\psi) =-\frac{GM_p r^2}{a^3}
 \left(\frac{4\pi}{5}\right) 
\sum^{m=2}_{m=-2}Y_{2,m}(\theta', \phi') Y_{2,m}(\pi/2, 0)\exp(-{\rm i}m \Phi),
\label{jpe0}
\end{equation}
where $a$ is the semi-major axis of the circular orbit  that lies in the plane 
$\theta'= {\rm \pi}/2$,  $P_2$ is the usual Legendre polynomial, $ Y_{2,m}(\theta', \phi'),$  denotes the usual spherical harmonic of degree $l'=2$
and $\cos\psi =\sin\theta'\cos(\Phi-\phi').$
The  azimuthal angle of the line joining the components is $\Phi.$
Both $\Phi$ and $\phi'$ are measured from  from the $X' $ axis.
 
 For the circular orbit we set $\Phi = n_ot +\varpi ,$
where   $n_o$  is the mean motion.
 The reference  angle, $\varpi,$  measured from the $X'$ axis can be taken to be the  longitude of the apsidal line
 once the orbit is slightly perturbed.  
 Following SP we  allow  $\varpi$  to vary with time on a time scale long compared to $n_0^{-1}$ as the orbit, being  influenced by tides
 is slightly non Keplerian. Thus the angle $\Phi$ increases at a mean rate given by
\begin{align}
\left\langle\frac{d\Phi}{dt}\right\rangle = n_0 + \left \langle\frac{d\varpi}{dt}\right \rangle,\label{epicy}
\end{align}
 where the enclosure in the angled brackets denotes a time average. Accordingly
 

\begin{equation}
	U=- 	\left(\frac{4\pi}{5}\right)  \frac{GM_p}{a^3} \, r^2
	\sum^{m=2}_{m=-2}Y_{2,m}(\theta', \phi') Y_{2,m}(\pi/2, 0) 
 \exp(-{\rm i}m \left[ n_o t+ \varpi\right]),
	\label{jp1F}
\end{equation}  


\subsection{ Expressing the perturbing potential in terms of spherical harmonics defined in the stellar frame}
We use Wigner matrices \citep[see e.g.][] {KMV} to express spherical harmonics defined in the orbit frame as a linear combination of spherical harmonics expressed in coordinates defined in the stellar frame. 
{ These are defined in the same way as for the orbit frame but with reference to the $(X,Y,Z)$ system.}
Thus
\begin{equation}
 Y_{2,m}(\theta', \phi')  = \sum_{n=-2}^{n =2}D^{(2)}_{n,m} ( 0,\beta,\gamma)Y_{2,n}(\theta, \phi),
\label{jpe5}
\end{equation}
where the coefficients or  Wigner matrix elements $D^{(2)}_{n,m}( 0,\beta,\gamma) $, 
being expressed using standard notation,
 depend on the  angles $\beta$ and $\gamma$ {defined in Section \ref{Geometry}, }which specify the  magnitudes of
 the angles of rotation required to transform the $(X,Y,Z)$ system to the $(X',Y',Z')$ system.
 First this requires  a rotation through an angle $\gamma$ about the $Z$ axis, followed by a rotation through an angle $\beta$ about the original $Y$ axis (see Section \ref{Geometry} and IP).  
 Also we may write
\begin{equation}
D^{(2)}_{n,m}( 0,\beta,\gamma) =  \exp(-{\rm i}m\gamma) \,  d^{(2)}_{n,m}(\beta) 
\label{jpe51a}
\end{equation}
where   $d^{(2)}_{n,m}$ is an element of
Wigner's (small) d-matrix  and is real   \citep[see e.g.][and IP] {KMV}.

\subsection{The inverse transformation} \label{invtrans}
The inverse transformation to  (\ref{jpe5}) expresses spherical harmonics
in the stellar frame in terms of those in the orbit frame.
The $(X',Y',Z')$ system converts to to the $(X,Y,Z)$ system
 through a rotation $-\beta$ about the $Y'$ axis  followed by a rotation through an angle $-\gamma$
about the $Z'$ axis.
Thus the transformation corresponding to the inverse of (\ref{jpe5}) is given by
\begin{equation}
 Y_{2,m}(\theta, \phi)  = \sum_{n=-2}^{n =2}D^{(2)}_{n,m} ( -\gamma,-\beta,0) \, Y_{2,n}(\theta', \phi'),
\label{jpe5inv}
\end{equation}
\begin{align}
&\hspace{-4.1cm}{\rm Corresponding}\hspace{2mm}  {\rm to} \hspace {2mm} {\rm(\ref{jpe51a})} \hspace{ 2mm} {\rm we} 
\hspace{2mm}{\rm  have}\hspace{2mm} D^{(2)}_{n,m}( -\gamma ,-\beta,0) =  \exp({\rm i}n\gamma) \, d^{(2)}_{m,n}(\beta). 
\label{jpe51ainv}
\end{align}
The inverse transformation (\ref{jpe5inv}) also follows  from  the big $D$ and small $d$ matrices being unitary.  
\begin{equation}
\hspace{-9cm}{\rm Thus} \hspace{2mm} D^{(2)}_{n,m}( -\gamma ,-\beta,0)=(D^{(2)}_{m,n}( 0,\beta,\gamma))^{*}\label {unitary}.
 \end{equation}

\subsection{The perturbing tidal potential in the stellar frame}\label{tidalpot}
\noindent Applying the Wigner transformation  (\ref{jpe5}) with (\ref{jpe51a}) 
 the perturbing tidal  potential  (\ref{jp1F}) 
is then defined with coordinates defined in the stellar frame.
One obtains  a sum of Fourier modes
 with  time dependence  through a factor, $\exp(-{\rm i}m n_o t ),$  in the  form
\begin{equation}
 U = \frac{r^2}{2} \sum_{m=-2}^{m=2}  {c}_{tid,m}\exp(-{\rm i}m\left(n_o t + \varpi+\gamma \right))
 \sum_{n=-2}^{n=2}d^{(2)}_{n,m}Y_{2,n}(\theta, \phi)
,\label{potpert}
\end{equation}
\begin{equation}
\hspace{-9.6cm} {\rm where}\hspace{2mm} c_{tid,m}  =-  \frac{8\pi GM_p}{5 a^3}  Y_{2,m}(\pi/2, 0).\hspace{4mm}
\label{jpe711e}
\end{equation}
We also note the following useful relations.
From the definition of spherical harmonics we have  $Y_{2,-m}(\theta, \phi) = (-1)^{m}Y^*_{2,m}(\theta, \phi) $ and the Wigner matrices satisfy
 $ D^{(2)}_{-n,-m}~=~(-1)^{(n+m)}( D^{(2)}_{n,m} )^*$ and  $ d^{(2)}_{-n,-m}~=~(-1)^{(n+m)}( d^{(2)}_{n,m} )^*.$ 
These relations are easily shown to imply that the sum in equation (\ref{potpert}) is real.
 Accordingly it also follows  from  equation (\ref{potpert})
    the primary´s tidal response can be  assembled by summing individual responses
   to the real parts of harmonically varying tidal potentials  of the form 
\begin{equation}
	U_{n,m} = r^2 c_{tid,m}\frac{d^{(2)}_{n,m}(\beta)}{(1+\delta_{m,0}\delta_{n,0})}Y_{2,n}(\theta, \phi) \, \exp(-{\rm i}m\left(n_o t + \varpi+\gamma \right)),
\end{equation}
with 
 $\delta_{i,j}$ denoting the Kronnecker delta.
 As the $\beta$ dependence is only through the factor $d^{2}_{n,m}$ we find it convenient to remove it from the response
 calculation by considering the tidal response to potentials of the form
 \begin{equation}
	{\cal U}_{n,m} =\frac{ r^2}{2} c_{tid,m}Y_{2,n}(\theta, \phi) \, \exp(-{\rm i}m\left(n_o t + \varpi+\gamma \right)) \label{potpert1}
\end{equation}
and then reintroduce it through applying a response  amplitude factor when required.  
Note  that, $-mn_o,$ is the forcing frequancy in the non rotating  frame  which corresponds
to the the forcing frequency, $\omega_{f,n,m} = n\Omega_s-mn_o,$ in the frame corotating with the star.

The Lagrangian displacement associated with the response to the perturbing potential ${\cal U}_{n,m}$ is $\mbox{{\boldmath$\xi$}}_{n,m}  \exp(-{\rm i}m\left(n_o t + \varpi+\gamma \right))$
as viewed in the non rotating frame. 
The associated Eulerian density perturbation  is  $\rho^\prime_{n,m} \exp(-{\rm i}m\left(n_o t + \varpi+\gamma \right))$ with similar expressions for the other perturbed state variables.

\subsection{The perturbation to the  external gravitational potential due to the primary  }\label{pertpot}
After separating out the exponential factor $\exp(-{\rm i}m\left(n_o t + \varpi+\gamma \right)),$ following PS
the perturbation to the external gravitational potential  produced by the forcing  potential $d^{(2)}_{n,m}(\beta){\cal U}_{n,m}$\footnote{An amplitude factor $d^{2}_{n,m}(\beta)$ is inserted at this point.}
  at position vector ${\bf R}$ 
in the stellar frame is
\begin{equation}
\psi_{n,m}' = -Gd^{(2}_{n,m}(\beta)\int_V \frac{\rho'_{n,m}({\bf r})}{| {\bf R} -{\bf r}|}dV,       
\end{equation} 
where the integral is taken over the volume of the primary.

For $|{\bf R}| >> |{\bf r}|$ we perform an expansion in inverse powers of $|{\bf R}|$ and spherical harmonics.
The dominant term then takes the form of a quadrupole as  (see PS)
\begin{equation}
\psi_{n,m}' = -\frac{4\pi G d^{2}_{n,m}(\beta)}{5R^3} 
Y_{2,n}(\theta,\phi) \int_V {\rho'_{n,m}({\bf r})}{r^2}Y_{2,n}^*(\theta,\phi) dV      
\end{equation} 
where $R = |{\bf R}|.$
  We may now assemble the response to the tidal potential given by
 equation (\ref{potpert}) using linear superposition with the result
 
 \begin{align}
&\hspace{-4mm}\psi'=\sum_{n,m} 
\psi_{n,m}'   \exp(-{\rm i}m\left(n_o t + \varpi+\gamma \right)) \equiv
 -\sum_{n,m}{\cal B}_{n,m}
 Y_{2,n}(\theta,\phi) \exp(-{\rm i}m\left(n_o t + \varpi+\gamma \right)) ,
 \label{extpotstar}      
\end{align}

\begin{align}
\hspace{-6.8cm} {\rm with}\hspace{3mm}  {\cal B}_{n,m} = \frac{4\pi G}{5R^3} 
 d^{(2)}_{n,m}(\beta)Q_{n,m},\hspace{3mm}\sum_{n,m}\equiv \sum_{n=-2}^{n=2}\sum_{m=-2}^{m=2} \hspace{2mm}{\rm and}\nonumber
\end{align}
\begin{align}
\hspace{-5.2cm} Q_{n,m}= \int_V {\rho'_{n,m}({\bf r})}{r^2}Y_{2,n}^*(\theta,\phi) dV\hspace{5mm} {\rm defining\hspace{1mm}  the \hspace{1mm} overlap \hspace{1mm} integral.}\nonumber
\end{align}
\noindent The component of the of the  specific torque in the $Z$ direction is then 
  \begin{align}
&T_z\equiv -\frac{\partial \psi'}{\partial \phi}=
\sum_{n,m}
{\rm i} n
 {\cal B}_{n,m}
 Y_{2,n}(\theta,\phi) \exp(-{\rm i}m\left(n_o t + \varpi+\gamma \right)) 
    \label{torqstar}  
\end{align} 

 \subsection{The potential perturbation in the orbit frame and its time average at the location of $M_p.$}\label{potaveMp}
 We make use of (\ref{jpe5inv}) - (\ref {unitary}) to transform the   spherical
 harmonics expressed  in terms of coordinates associated with the stellar frame that appear in equation (\ref{extpotstar} )
 to a linear combination of spherical harmonics 
 expressed in terms of coordinates defined in the  orbit frame. Equation (\ref{extpotstar} ) specifying  
 $\psi'$ then becomes

 \begin{align}
&\psi'= -
 \sum_{n,m,n'} 
 {\cal B}_{n,m}
  \, d^{(2)}_{n,n^\prime}(\beta) 
Y_{2,n'}(\theta', \phi')  \exp(-{\rm i}(m\left(n_o t + \varpi+\gamma ) -n'\gamma\right))
  \label{potpertorbit}
\end{align} 
\begin{align}
\hspace{-8.3cm}{\rm where} \hspace{1mm} {\rm we} \hspace{1mm}{\rm denote }\hspace{1mm} 
 \sum_{n,m,n'} \equiv \sum^{n=2}_{n=-2}
 \sum_{m=-2}^{m=2} \sum_{n'=-2}^{n' =2} .
\end{align}

 Setting $\theta'=\pi/2,$  $\phi' = \Phi$ and  $R = a$ 
 in equation(\ref{potpertorbit}) we obtain the potential perturbation
at the location of the companion as
 \begin{align}
&\psi'=
  -\sum_{n,m,n'}  
  {\cal{B}}_{n,m} 
  \,d^{(2)}_{n,n'}( \beta )Y_{2,n'}(\pi/2, 0)
 \exp({\rm i}\left (n'-m)(n_0t+\gamma+\varpi)\right),  
\end{align} 

Neglecting any possible slow variation of $\varpi,$ the time average of, $\psi',$ 
evaluated at $M_p$ found by integrating around the orbit is 
 \begin{align}
&\langle \psi' \rangle=
  -\sum_{n,m}   {\cal{B}}_{n,m}d^{(2)}_{n,m}( \beta )Y_{2,m}(\pi/2, 0)
\end{align} 
\noindent Noting the connection between $T_z$ and $\psi'$ indicated by equation(\ref{torqstar}),  
 the time average of $T_z$ is 
 given by
 \begin{align}
&\langle T_z\rangle =
 \sum_{n, m } {\rm i}n  {\cal{B}}_{n,m} \,d^{(2)}_{n,m}( \beta )Y_{2,m}(\pi/2, 0)
\end{align}

\subsection{The force per unit mass acting on $M_p$}\label{fperunitMp}
We write the  components of the force ${{\boldsymbol{{\cal{F}}}} }$ acting on $M_p$ resulting from $\psi'$ 
in the orbit frame as $({\cal F}_R,{\cal F}_{\Phi},{\cal  F}_{\perp}).$
These  act in the radial direction, the azimuthal direction and the direction perpendicular to the
orbital plane or $Z'$ direction respectively.
We have 
\begin{align}
\hspace{0mm} ({\cal F}_R,{\cal F}_{\Phi},{\cal F}_{\perp}) =
\bigg (-M_p\frac{\partial \psi'}{\partial R }\quad ,\quad -\frac{M_p}{R\sin\theta'}\frac{\partial \psi'}{\partial \phi'} \quad ,\quad \frac{M_p}{R}\frac{\partial \psi'}{\partial\theta'}  \bigg)\bigg{ |}_{\theta'=\pi/2,\phi'=\Phi}
\end{align}
These are readily obtained from equation(\ref{potpertorbit}) from which we obtain
\begin{align}
&\hspace{-3mm} {\cal F}_R= 
- \frac{3M_p}{a} \sum_{n,m,n'} 
 {\cal B}_{n,m}
  d^{(2)}_{n,n'}( \beta) \,Y_{2,n'}(\pi/2, 0)
   \exp({\rm i}\left (n'-m)(n_0t+\gamma+\varpi)\right),  
  \label{FR}
\end{align}

\begin{align}
&{\cal F}_{\Phi}= 
 \frac{{\rm i}M_p}{a} \sum_{n,m,n'} n' 
 {\cal B}_{n,m}
  d^{(2)}_{n,n'}( \beta) \,Y_{2,n'}(\pi/2, 0)
   \exp({\rm i}\left (n'-m)(n_0t+\gamma+\varpi)\right),  
  \label{FPHI}
\end{align} 
 
 \begin{align}
&\hspace{-3mm} {\cal F}_{\perp}= 
 -\frac{M_p}{a} \sum_{n,m,n'} 
 {\cal B}_{n,m}
  d^{(2)}_{n,n'}( \beta) \,Y^{'}_{2,n'}(\pi/2, 0)
   \exp({\rm i}\left (n'-m)(n_0t+\gamma+\varpi)\right),  
  \label{FPERP}
\end{align} 
where $Y_{2,n'}'(\theta'\phi') = \partial Y_{2,n'}(\theta' ,\phi')/\partial\theta'.$

\subsection{Gauss equation  for the evolution of the semi-major axis}\label{Gaussa}
The Gauss equations express the rate of change of orbital elements
in terms of the components of  the force per unit mass ${\bf F},$
assumed to act on $M_p.$ Because our coordinate system is a non inertial 
accelerating reference frame
with origin at the centre of mass  of the primary, and the force acting on $M_p$
produces an equal and opposite reaction acting on the primary, the force per unit mass, ${\bf F},$ determining the evolution of the orbit
is obtained  from ${\boldsymbol{{\cal F}}}$ by dividing by the reduced mass, $M_pM_*/(M_p+M_*)$
rather than $M_p.$ For a solar mass star with a planetary mass companion this correction, though implemented,
is negligible.

The Gauss equation governing the rate of change of the semi-major axis  is
\begin{align}
\frac{da}{dt}=\frac{2a^{3/2}}{\sqrt{G(M_p+M_*)(1-e^2)}}\left(F_R \, e\sin(\Phi-\varpi) +F_{\Phi}(1+e\cos(\Phi-\varpi))\right),\label{adotG}
\end{align} 
where, $e,$ is the orbital eccentricity which we set to  zero  for circular orbits.
Setting $e=0$ in (\ref{adotG}),    noting  the above remarks  about determining ${\bf F}$, making use of equation ({\ref{FPHI}), and then taking the time average,  we obtain
  
  \begin{align}
&\left \langle\frac{da}{dt}\right\rangle =\frac{2 a^{2}}{GM_*} 
\sum_{n,m} {\rm i}mn_o  {\cal{B}}_{n,m}
  d^{(2)}_{n,m}
  Y_{2,m}(\pi/2, 0),
  \label{adot4}           
  \end{align}
 Using the form of ${\cal B}_{n,m}$ given by (\ref{extpotstar}) together with the relations given in Section \ref{tidalpot}
and  noting that the overlap integral is 
such that,  $Q_{-n,-m}^{*} = (-1)^{m} Q_{n,m},$ 
it follows that the expression (\ref{adot4}) is real and can be written in the form
 \begin{align}
	\left\langle \frac{da}{dt}\right\rangle= -\frac{32\pi  n_0}{5M_*a} \im \left[ Y_{2,2}(\pi/2,0)
	\sum^{j=2}_{j=-2}\left(d^{(2)}_{j,2}\right)^2Q_{j,2}
	\right]
	\label{adotavecirc}
\end{align}
We remark that because the Wigner matrix with real elements, $d^{2}_{n,m},$ is unitary
the sums ${\big. \Sigma}_{j=-2}^{j=2}  d^{(2)}_{j,n}d^{(2)}_{j,n} $ are equal to unity for any possible $n.$ This means that for a spherical non rotating star
for which $Q_{j,2}$ does not depend on $j,$ $ \langle da/dt\rangle$ does not depend on $\beta$ as expected.

 \begin{figure}
\hspace{-0.3cm}	\includegraphics[height=10cm, width=12cm]{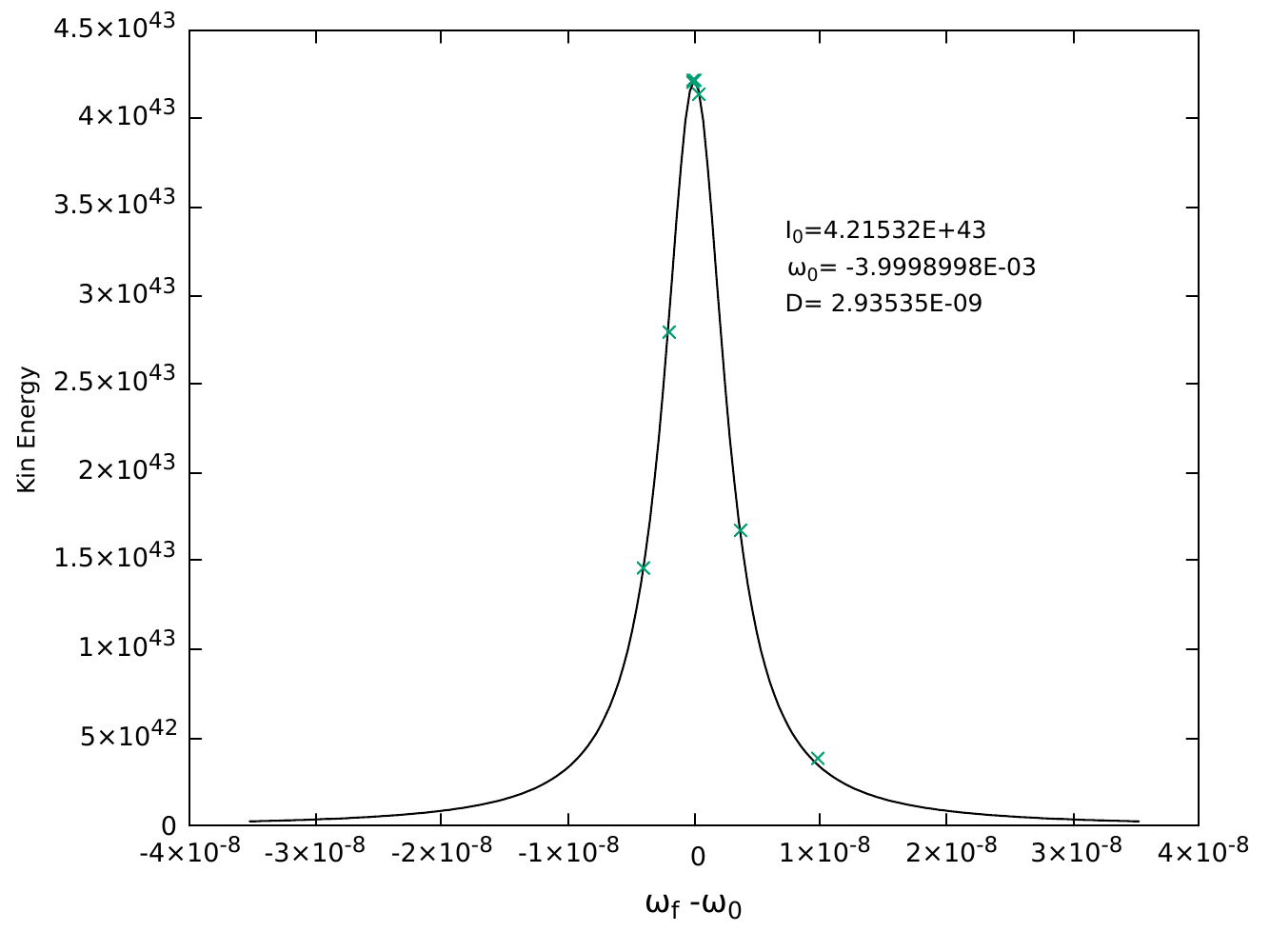}

	\includegraphics[height=10cm,width=12cm]{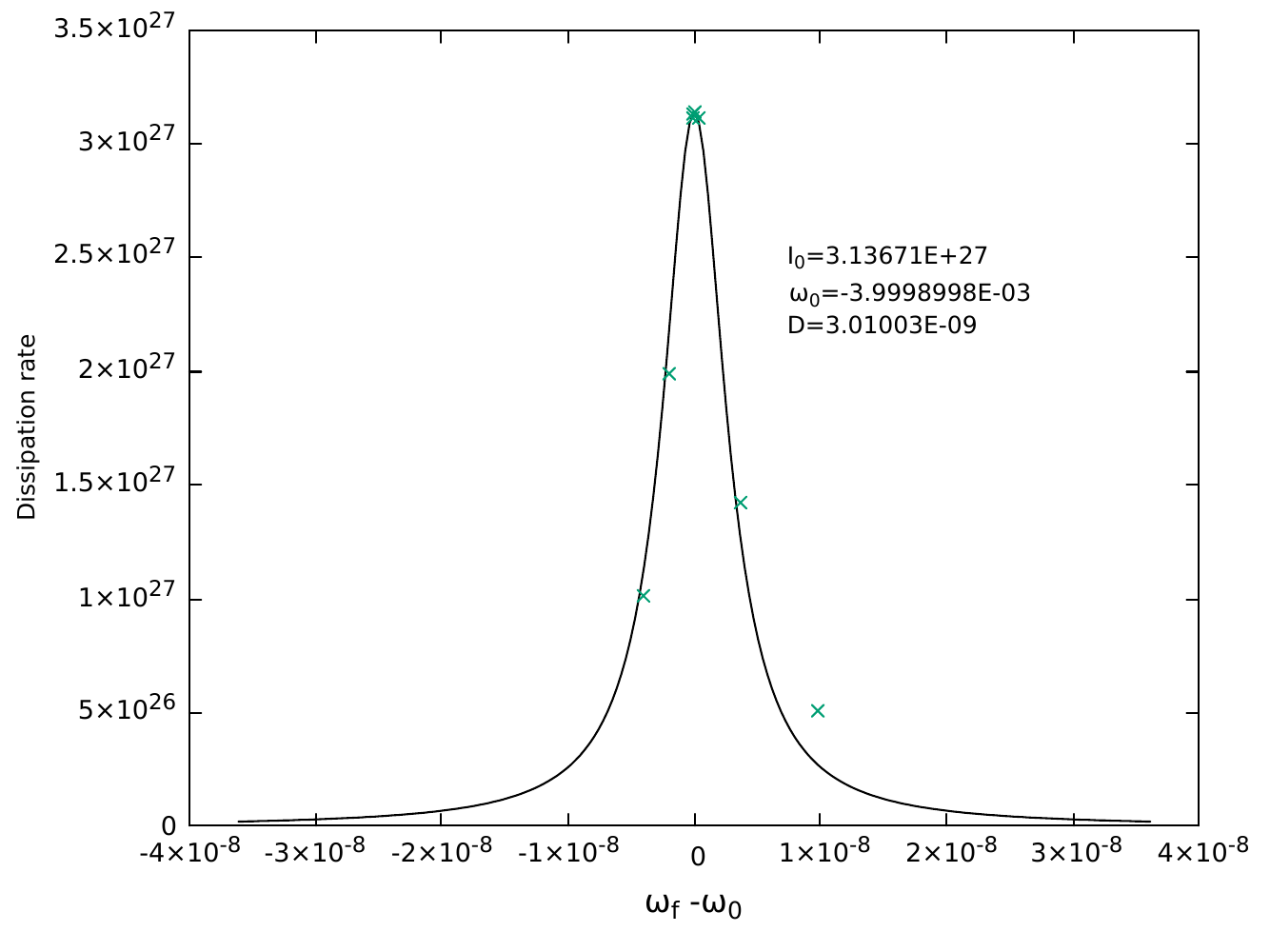}
	\caption{The { upper}  panel shows the  resonance curve (with resonance frequency $\omega_0$ and full resonance half width, $D$) associated with the kinetic energy response (erg) for the $n_r=0$ $r$-mode with $n=-1$ and $m=-2.$ The { lower}  panel shows the corresponding resonance curve obtained for the viscous dissipation rate  (erg/s). Note that the subscripts $n,$ and $m$ have been removed from
		the forcing frequency $\omega_f.$ }  \label{Disp-3107P2}
\end{figure}

\subsection{Gauss equation  for the evolution of the  orbital inclination}\label{Gaussinc}
The Gauss equation for the rate of change of the inclination is
\begin{align}
&\frac{d i }{dt}=
\frac{F_{\perp}R\sin(\Phi+\gamma)}{\sqrt{G(M_p+M_*)a(1-e^2)}}\label{Gaussequation4}
\end{align}
We recall that the orbital inclination $i$ is the angle between the orbital angular momentum $\bf{L}$ and the total angular momentum $\bf{J}$ of the system.
Substituting $F_{\perp},$ which is given by  equation (\ref{FPERP}) into equation (\ref{Gaussequation4}), taking into account the comments at the beginning of 
Section \ref{Gaussa},  and  setting  $R=a,$  we obtain 
\begin{align}
&\frac{d i }{dt}=
-\frac{n_o a}{GM_*} \sum_{n,m,n'} 
 {\cal B}_{n,m}
  d^{(2)}_{n,n'}( \beta)Y^{'}_{2,n'}(\pi/2, 0)\sin(\Phi+\gamma)
   \exp({\rm i}\left (n'-m)(n_0t+\gamma+\varpi)\right)
\label{Gaussequation4.1}
\end{align}
Noting that $\Phi=n_ot+\varpi$ and taking the time average we  obtain 
\begin{align}
&\bigg\langle \frac{d i }{dt}\bigg\rangle =
-\frac{n_o a}{2{\rm i}GM_*} 
\sum_{n,m} 
 {\cal B}_{n,m}
 ( d^{(2)}_{n,m-1}( \beta)Y^{'}_{2,m-1}(\pi/2, 0)
 -d^2_{n,m+1}(\beta)Y^{'}_{2,m+1}(\pi/2,0))
\label{Gaussequation4.2}
\end{align}
Using the expressions given with equation (\ref{extpotstar}) to express the result in terms of the overlap integrals and performing the sum over $m,$
we find
\begin{align}
	&\bigg\langle \frac{d i }{dt}\bigg\rangle =
	-\frac{4\pi  n_o}{5a^{2} M_*} 
	\im\left[Y^{'}_{2,1}(\pi/2, 0)
	\sum^{j=2}_{j=-2} (d^{(2)}_{j,1}d^{(2)}_{j,2}Q_{j,2}-d^{(2)}_{j,1}d^{(2)}_{j,0}Q_{j,0})
	\right]
	\label{Gaussequation4.3x0}
\end{align}
In this case we note that because the Wigner matrix with real elements, $d^{2}_{n,m},$ is unitary
the sums ${\big. \Sigma}_{j=-2}^{j=2}  d^{(2)}_{j,n}d^{(2)}_{j,m} $ are zero for $n\ne m.$ This means that for a spherical non rotating star
for which $Q_{j,n}$ does not depend on $j,$  for any possible $n,$ we have $ \langle di/dt\rangle=0$ as expected.

\begin{figure}
	\hspace{-1.0cm}	
	\includegraphics[angle=0,width=1\columnwidth]{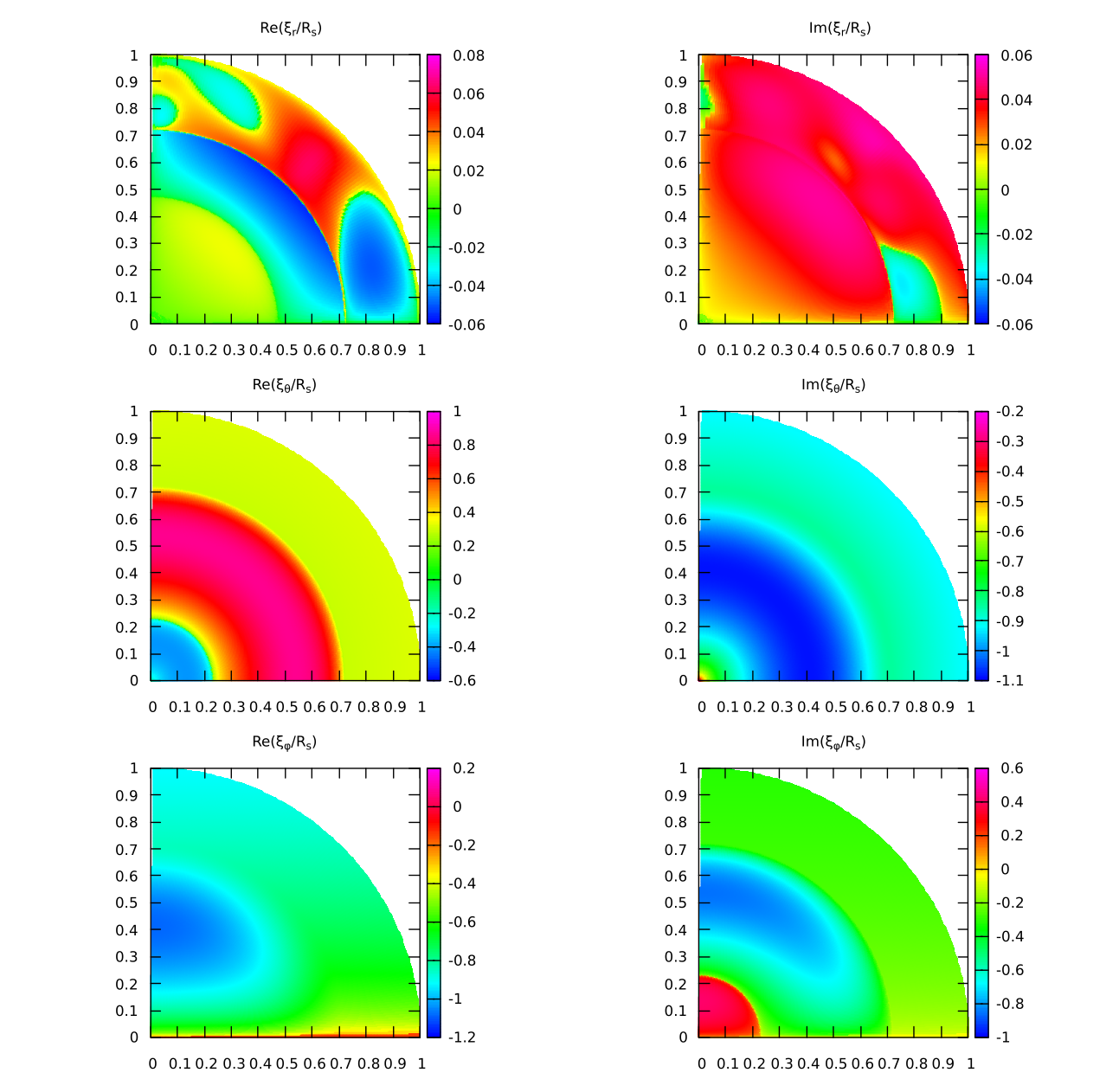}.
	\caption{Contour plots  in the primary's meridional plane $\phi=0$ for the  dominant  $l'$=1 r mode,
	  at the  resonance frequency $\omega_0=-3.999899{ 831} \times 10^{-3}$, this being the closest one to $-\Omega_s.$
		These were obtained by resonant forcing with, ${\cal U}_{-1,-2},$  with $(n, m)=(-1,-2) ,$  and $\Omega_s=4.0\times10^{-3}\Omega_{c}$ { (see Fig. \ref{Disp-3107P2}).}
		The forcing frequency  $\omega_{f,-2,-1}= 2 n_o-\Omega_s \equiv \omega_0$ is very close to $-\Omega_s$ so that $n_0\ll\Omega_s.$
		The Cartesian coordinates along the two axes indicate the relative radius $r/R_*$. 
		The vertical colour bars on the right indicate the local value of  sign($|\xi_x|^{\frac{1}{4}},\xi_x)$, where  $\xi_x$ is  the  component of the displacement vector illustrated
		in units of $R_s$.\label{c13.3107P2} }
\end{figure} 

\clearpage

 	\begin{figure}
	\vspace{-0.0cm}
\hspace{0.3cm}	\includegraphics[height=10cm, width=12cm]{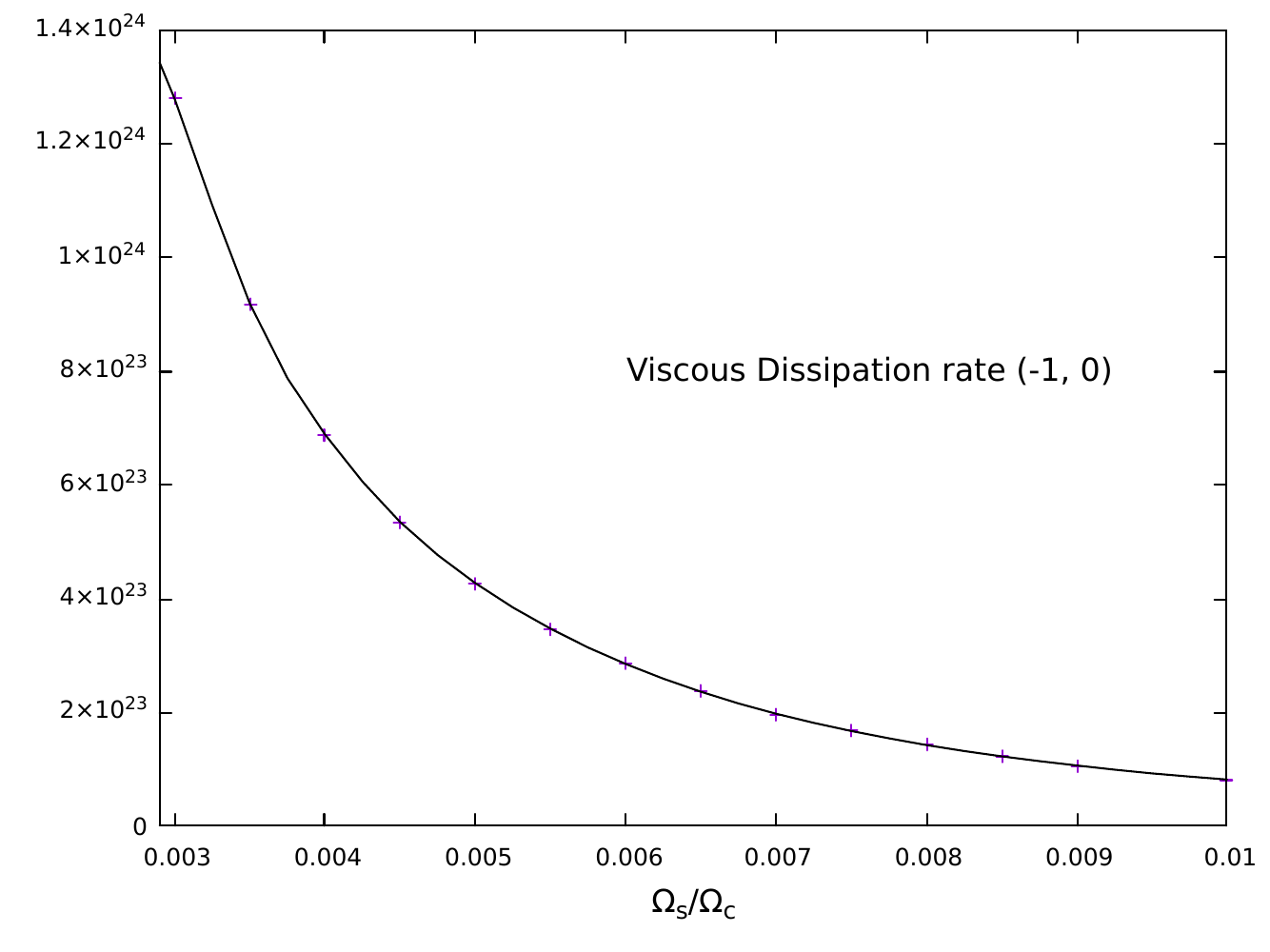}

	\vspace{-0.1cm}
\hspace{0.3cm}	\includegraphics[height=10cm, width=12cm]{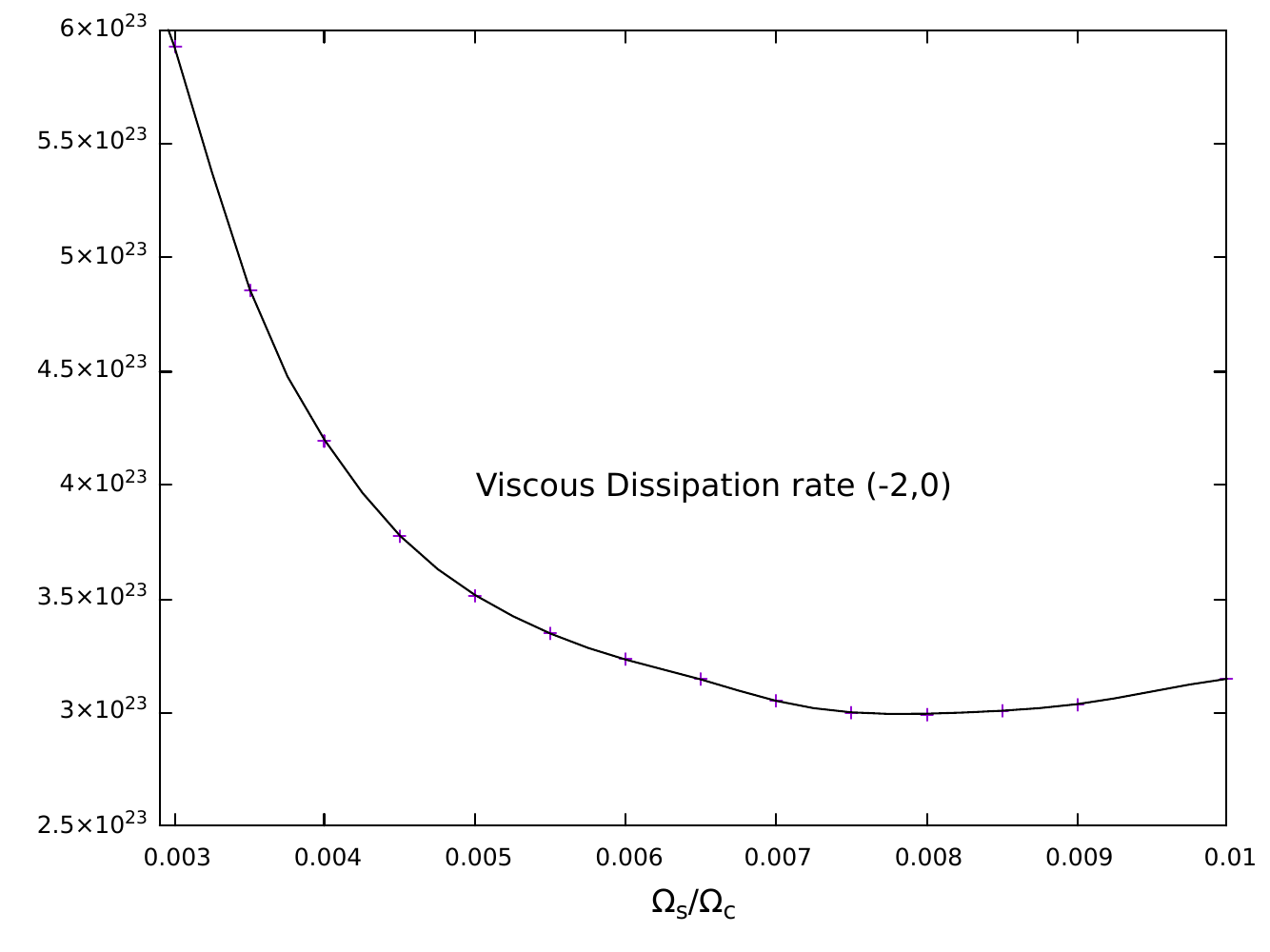}
	\caption{The { upper} panel shows the viscous dissipation rate in the convective envelope, 
	 $ -\dot{E}_{kin,-1,0} \equiv -\dot{E}_{kin,1,0}$ (erg/s),  for $ (n,m)= (-1,0)$  forcing  as a function of the stellar spin rate. 
	The {lower}  panel shows the same but for for forcing with $(n,m)=(-2,0)$.} \label{Disprate1}
\end{figure}

\begin{figure}
	\includegraphics[height=10cm, width=12cm]{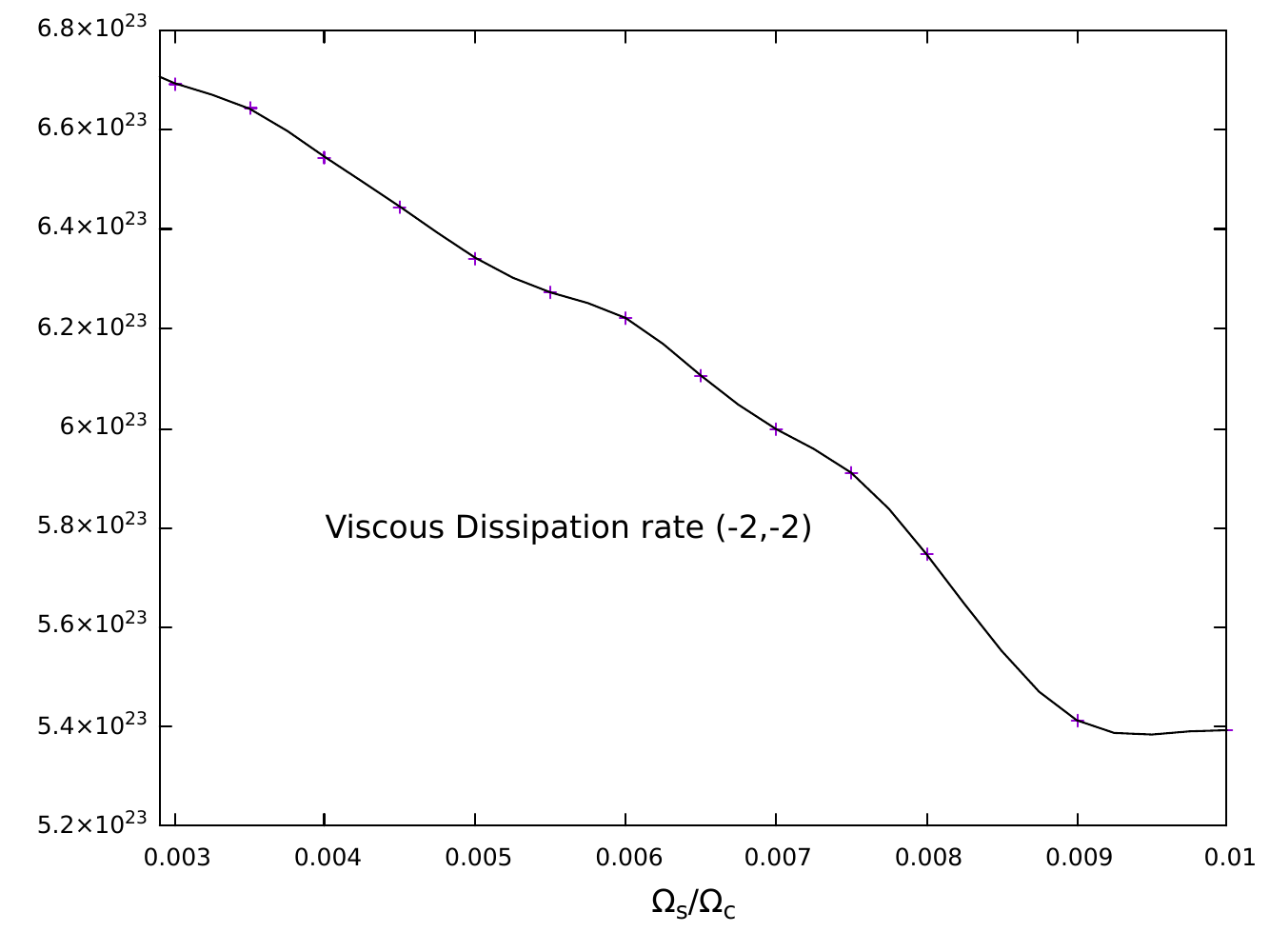}
	\caption{The viscous dissipation rate $ -\dot{E}_{kin,-2,-2} \equiv -\dot{E}_{kin,2,2}$ (erg/s) in the convective envelope for $(n,m)= (-2,-2)$
	 forcing  as a function of the stellar spin rate.} \label{Disprate2}
\end{figure}

\section{Discussion of the evolution equations for semi-major axis and inclination}\label{DiscGausseq}
Following the discussion in Section 5.4 of PS,  we note that
the overlap integrals appearing in equations (\ref{adotavecirc}) and (\ref{Gaussequation4.3x0}) can be related
 to the mean rate of change of the  kinetic energy associated with 
forcing due to  the real part of the  corresponding potential given by (\ref{potpert1}).
This relation is given by  a minor adaption of equation (30) of PS \footnote{Note that this expression is a factor of $2$ smaller than expected
 from equation (30) pf PS. This is because the normalisation of the corresponding
potential (12) is a factor of $2$ smaller. That is because  contributions with $m=-2$ and $m=2$ have been treated separately here.} 
  in the form
\begin{align}
\frac{dE_{kin,n,m}}{dt} = -\frac{c_{tid,m} \omega_{f,n,m}}{4} Im (Q_{n,m}). \label{Edot}
 \end{align}
 This quantity  is expected to be negative definite.
 
 Substituting into equation (\ref{adotavecirc}) we obtain
  \begin{align}
	\left\langle\frac{da}{dt}\right\rangle = -\frac{16  n_0 a^2}{GM_*M_p}  
	 \sum^{j=2}_{j=-2}\frac{\left(d^{(2)}_{j,2}\right)^2(dE_{kin, j,2}/dt)}{\omega_{f,j,2}} .
	\label{adotaveedot}
\end{align}
Here we recall that $\omega_{f,j,m}= -mn_o+j\Omega_s.$ 
Similarly from equation (\ref{Gaussequation4.3x0}) governing the mean rate of change
of inclination we obtain
\begin{align}
	&\bigg\langle \frac{d i }{dt}\bigg\rangle =
	-\frac{2  n_o a
	 Y^{'}_{2,1}(\pi/2, 0)}{{ GM_pM_*} }
	\sum^{j=2}_{j=-2} \left(\frac{d^{(2)}_{j,1}d^{(2)}_{j,2} (dE_{kin, j,2}/dt)}{\omega_{f,j,2}Y_{2,2}(\pi/2, 0)} 
	 -\frac{d^{(2)}_{j,1}d^{(2)}_{j,0}(dE_{kin.j,0}/dt)}{\omega_{f,j,0}Y_{2,0}(\pi/2, 0)}    \right)
	\label{Gaussequation4.3x}
\end{align}
We note that the $\beta$ dependence in equations (\ref{adotaveedot}) and (\ref{Gaussequation4.3x}) is contained entirely within the
Wigner small $d$  matrix elements. In the  case of, $a,$  the rate of evolution is determined by the five rate of kinetic energy changes  $(dE_{kin.,j,2}/dt)$ 
for $j=-2,-1,0,1,2.$ In the case of the inclination, the evolution also depends on these together with  $ (dE_{kin, j,0}/dt)$ for $j=1$ and $j=2,$ Contributions from negative values of $j$
are related to these and 
in this case there is no contribution from  $j=0.$   
\subsection{The relation between $\langle d i/dt\rangle$ and $\langle d\beta/dt\rangle$}
The angles $i$ and $\beta$ are related by the conservation of the total angular momentum, ${\bf J}={\bf L}+{\bf S}.$
Setting $J= |{\bf J}|,  L= |{\bf L}|,  S= |{\bf S}|,$ and considering the components of ${\bf J}$, perpendicular and parallel to {\bf L},  we have (see IP)
\begin{align}
J\cos i = L + S\cos\beta, \hspace{2mm} {\rm and}\hspace{2mm} J\sin i = S\sin\beta.
\end{align}
Differentiating the above expressions with respect to time while using the fact that ${\bf J} $ is conserved, making use of (\ref{adotaveedot}),  (\ref{Gaussequation4.3x})
 and substituting 
the required values of $Y_{2,2}, Y_{2,0} $ and $Y'_{2,1}$  we obtain
\begin{align}
\left\langle\frac{d\beta}{dt}\right\rangle = \left(1+\frac{L}{S}\cos\beta\right)\left\langle\frac{di}{dt}\right\rangle +\frac{\sin\beta}{S}\left\langle\frac{dL}{dt} \right\rangle
\end{align}
\begin{align}
	\left\langle \frac{d \beta }{dt}\right\rangle  =&
	-\frac{4  n_o  a \left(1+L\cos\beta/S\right)
	 }{G M_pM_*}
	\sum^{j=2}_{j=-2} \left(\frac{d^{(2)}_{j,1}d^{(2)}_{j,2} (dE_{kin, j,2}/dt)}{\omega_{f,j,2}}
	 +\sqrt{\frac{3}{2}}\frac{d^{(2)}_{j,1}d^{(2)}_{j,0}(dE_{kin, j,0}/dt)}{\omega_{f,j,0}}
	   \right)\nonumber\\
	&-\frac{8\sin\beta}{S} 
	 \sum^{j=2}_{j=-2}\frac{\left(d^{(2)}_{j,2}\right)^2(dE_{kin, j,2}/dt)}{\omega_{f,j,2}}
	\label{Gaussequation4.3y}
\end{align}
{\textcolor{red}{\subsection{Estimating  $\langle d\beta/dt\rangle$ when  $n_o/\Omega_s$ is large}\label{estbetadot}
We are primarily interested in the situation when the orbital period is significantly 
shorter than the rotation period such as for a hot Jupiter with an orbital period of a few days
and a solar type star with rotation period an order of magnitude larger.}

\textcolor{red}{In this case we suppose that the dominant contribution to the righthand side of equation (\ref{Gaussequation4.3y})  comes from the forcing terms 
with $m=0$ and hence  zero forcing frequency, $j=\pm1,$
and $j=\pm2.$ We remark that zero frequency forcing with $j=0$ corresponds to the production of a static tide which is not expected to result in orbital evolution.}

\textcolor{red}Forcing  with $m=\pm 2$ occurs with an associated  forcing frequency a factor $\sim n_o/\Omega_s$ larger in magnitude than is the case for $m=0.$
Thus  contributions from such terms  are  expected to be smaller than those arising from $m=0$ by at least this factor given  comparable energy dissipation rates.
In addition the terms on the right hand side of (\ref{Gaussequation4.3y}) $\propto d^{(2)}_{j,1}d^{(2)}_{j,2}$  cancel when summed over $j$
in the limit $\Omega_s\rightarrow 0.$ Thus these terms, derived from  $m= \pm 2$ forcing, are expected to be smaller in magnitude  than terms derived from  zero frequency 
forcing by a factor $\sim (n_o /\Omega_s)^2$
for the comparable energy dissipation rates we find. }

\textcolor{red}{ Thus we neglect contributions arising from terms with $m=\pm 2$
and write
\begin{align}
	\left\langle \frac{d \beta }{dt}\right\rangle  =&
	-\frac{4  n_o  a \left(1+L\cos\beta/S\right)
	 }{G M_pM_*}\sqrt{\frac{3}{2}}
	\sum^{j=2}_{j=-2,\ne 0} 
	\frac{d^{(2)}_{j,1}d^{(2)}_{j,0}(dE_{kin, j,0}/dt)}{\omega_{f,j,0}}.
	\label{Gaussequation4.3z}
\end{align}
In this case by making use of (\ref{Edot}) and noting that $Q_{-j,0}=Q_{j,0}^{*},$  and\\  $dE_{kin.-n,-m}/dt =dE_{kin. n,m}/dt ,$ we can write (\ref{Gaussequation4.3z}) as
\begin{align}
	&\left\langle \frac{d \beta }{dt}\right\rangle  =
	-\frac{4  n_o  a \left(1+L\cos\beta/S\right)
	 }{G M_pM_*}\sqrt{\frac{3}{2}}\times\nonumber\\
	&\left(\frac{(d^{(2)}_{1,1}d^{(2)}_{1,0}-     d^{(2)}_{-1,1}d^{(2)}_{-1,0} )(dE_{kin.1,0}/dt)}{\omega_{f,1,0}}+
	\frac{(d^{(2)}_{2,1}d^{(2)}_{2,0}-     d^{(2)}_{-2,1}d^{(2)}_{-2,0} )(dE_{kin.2,0}/dt)}{\omega_{f,2,0}}	\right)
	\label{Gaussequation4.3z1}
\end{align}
Making use of  the small $d$ Wigner matrix elements given in appendix \ref{Wigner}, we can write this as
\begin{align}
	\left\langle \frac{d \beta }{dt}\right\rangle  =&
	\frac{6  n_o  a \left(1+L\cos\beta/S\right)\sin\beta
	 }{G M_pM_*\Omega_s}\left( \cos^2\beta
	 \frac{dE_{kin, 1,0}}{dt}+\frac{1}{4}\sin^2 \beta \frac{dE_{kin, 2,0}}{dt}\right)
	\label{Gaussequation4.3z2}
\end{align}
This compares with $a^{-1} \langle da/dt \rangle$ evaluated in the limit $\beta=0$ by making use of  equation (\ref{adotaveedot}).
We find 
 \begin{align}
	\frac{1}{a}\left\langle\frac{da}{dt}\right\rangle = \frac{8  n_0 a}{GM_*M_p}  
	 \frac{(dE_{kin, 2,2}/dt)}{(n_o-\Omega_s)} ,
	\label{adotaveedot1}
\end{align}
which  also applies in the limit $\Omega_s\rightarrow 0.$
Equations (\ref{Gaussequation4.3z2}) and (\ref{adotaveedot1} give two equations for $\beta$ and $a,$ or equivalently $L.$
These must be complemented by making use of the expression for the square of the magnitude of the total angular momentum 
which has been  assumed to be conserved in the form
$J^2=L^2+S^2+2LS\cos\beta = (S+L\cos\beta)^2+L^2\sin^2\beta.$ This can be used to specify ,$S,$ and so complete the system. }

\textcolor{red}{We may write 
\begin{align}
\frac{L}{S}= \frac{M_pM_*}{M_p+M_*}\frac{\sqrt{G(M_p+M_*)a}}{I\Omega_s}\approx
 \frac{M_p}{M_*}\frac{\Omega_c}{k\Omega_s}\left(\frac{\Omega_c}{n_o}\right)^{1/3},\label{LOS}
\end{align}
where the moment of inertia $I=kM_*R_*^2$ and we have assumed $M_p/M_*\ll 1.$
Making use of (\ref{LOS}) equation  (\ref{Gaussequation4.3z2}) can be written
\begin{align}
	\hspace{-3mm}\left\langle \frac{d \beta }{dt}\right\rangle =&
	 \frac{6n_o^{1/3}\Omega_c^{2/3}R_*}{\Omega_s GM_pM_*}
	\left(1+ \frac{M_p}{M_*}\frac{\Omega_c}{k\Omega_s}\left(\frac{\Omega_c}{n_o}\right)^{1/3}\cos\beta \right)\sin\beta\left(\frac{dE_{kin, 1,0}}{dt}\cos^2\beta
	 +\frac{\sin^2 \beta}{4} \frac{dE_{kin, 2,0}}{dt}\right).
	\label{Gaussequation4.3z3}
\end{align}
Specialising to the case when the orbital angular momentum significantly exceeds the spin angular momentum
which is for the most part appropriate here, only the second term in the first set of brackets in (\ref{Gaussequation4.3z3}) need be retained. Thus we have.
\begin{align}
	\left\langle \frac{d \beta }{dt}\right\rangle  =&
	6  \left( \frac{R_*}{kGM_*^2}\right)
	\left(\frac{\Omega_c}{\Omega_s}\right)^2 \cos\beta\sin\beta\left(\frac{dE_{kin, 1,0}}{dt}\cos^2\beta
	 +\frac{\sin^2 \beta}{4} \frac{dE_{kin, 2,0}}{dt}\right)
	\label{Gaussequation4.3z30}
\end{align}
We can compare the time scales for the evolution of the  
 semi-major axis and the inclination angle by considering
\begin{align}
	\frac{\left\langle d \beta /dt\right\rangle }{  a^{-1} \langle da/dt \rangle} =&
	  \left( \frac{3M_pR_*(n_o-\Omega_s)}{\hspace{-10mm}4M_* a\hspace{1mm} k\hspace{1mm} n_o}\right)
	\left(\frac{\Omega_c}{\Omega_s}\right)^2 \cos\beta\sin\beta
	\frac{\left(\frac{dE_{kin, 1,0}}{dt}\cos^2\beta
	 +\frac{\sin^2 \beta}{4} \frac{dE_{kin, 2,0}}{dt}\right)}
	{  \frac{ dE_{kin, 2,2}}{dt}  }.
	\label{Gaussequation4.3z3a}
\end{align}
\subsubsection{Dependence on $\beta$ and an estimate of tidal evolution time scales.}\label{testimate}
From equation (\ref{Gaussequation4.3z30}) $\beta$ has stationary points at $\beta=0, \pi/2,$ and $\pi.$
The first and the last of these are stable while the one corresponding to a polar orbit is unstable.
Even so evolution in its neighbourhood will be slowed down suggesting a possible accumulation of systems
with polar orbits if partially effective tides operate on a uniform initial distribution of misalignments
(see Section \ref{Discuss} below).  Note that if we take into account the magnitude of the spin angular momentum 
in comparison to the orbital angular momentum and use   
 equation (\ref{Gaussequation4.3z3}) the unstable point is shifted to be slightly in excess of $\pi/2$
 when this is small.
 }

\textcolor{red}{If the energy dissipation rates are comparable as found here \footnote{For quasi-static { responses (in the limit of zero frequency)}  of a spherical  star
$dE_{kin, 1,0}/dt= dE_{kin, 2,0}/dt = (2/3)dE_{kin, 2,2}/dt$}
    and $n_o\gg \Omega_s$ then, in the limit of small $\beta$,  
we have the simple result
\begin{align}
	\frac{\beta^{-1}\left\langle d \beta /dt\right\rangle }{  a^{-1} \langle da/dt \rangle}\approx &
	  \left( \frac{3M_pR_*}{\hspace{-0mm}4M_* a\hspace{1mm} k\hspace{1mm}}\right)
	\left(\frac{\Omega_c}{\Omega_s}\right)^2 
	\label{Gaussequation4.3z3b}
\end{align}
For $\Omega_s/\Omega_c= 0.004,  a/R_s=10,  k=0.08,$ and $M_p/M_*=0.001,$ 
 This yields a ratio of  $59.$ Thus we expect the inclination angle to   
 evolve about $60$ time faster than the semi-major axis for small $\beta.$
On the other hand  Equation (\ref {adotaveedot1}) gives  $|a (\langle da/dt \rangle)^{-1} | \approx 2.5\times 10^{12}(|dE_{kin, 2,2}/dt|/6\times10^{23}cgs)^{-1} y. $ }

\textcolor{red}{From the results presented below ( see also PS) a characteristic value of   $|dE_{kin, 2,2}/dt)| \sim 6\times10^{23} cgs.$
This indicates that $|\beta(\langle d\beta/dt\rangle )^{-1}| \sim 4\times 10^{10}y$ suggesting that for the parameters adopted the  evolution of $\beta$ will be 
be modest  over the lifetime of a solar type star (but see Section \ref{incMpef}  below for the scaling with planet mass and orbital period ) .}
\textcolor{blue}{ From the above considerations we see  that the time scale for, $a,$ to change is expected to be $\sim 2\times10^{12} y.$  Thus only a small change
to the orbital period is expected during the main sequence lifetime.}

\begin{figure}
	\hspace{-1.0cm}	
	\includegraphics[angle=0,width=1\columnwidth]{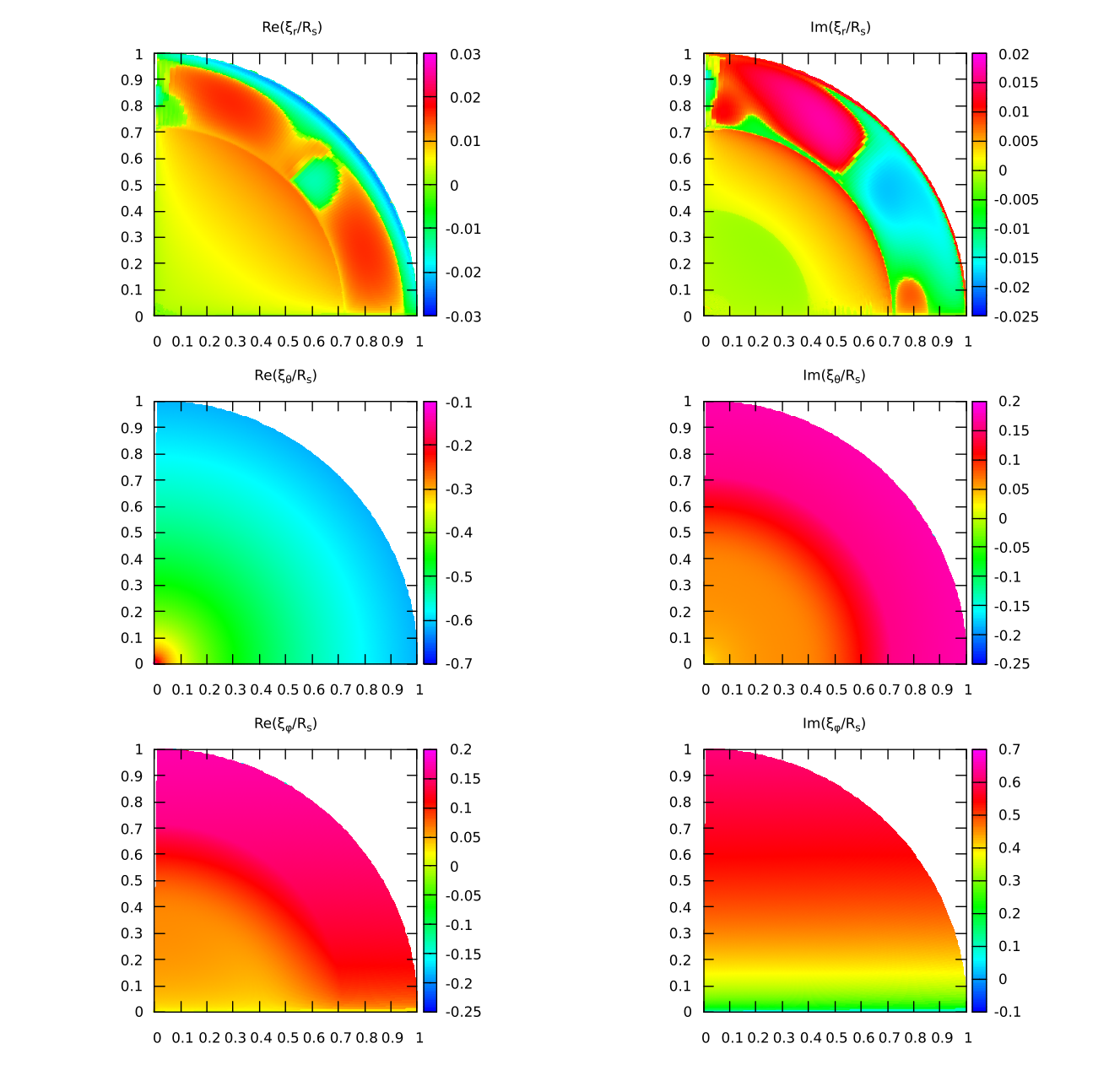}.
	\caption{Contour plots  in the primary's meridional plane $\phi=0$ 
		showing the response to forcing with, ${\cal U}_{-1,0}, (n,m)= (-1,0)),$  for $\Omega_s=3.0\times10^{-3}\Omega_{c}.$
		{ For these values of $n$ and $m$ } the forcing frequency is $-\Omega_s$  corresponding to a time independent perturbation in the non rotating frame. 
		{This calculation}  was done for $a/R_*=10.$ The response for other values can be obtained by applying the scaling factor $(a/R_*)^3.$ 
		The Cartesian coordinates along the two axes indicate the relative radius $r/R_*$. 
		The vertical colour bars on the right indicate the local value of  sign($|\xi_x|^{\frac{1}{4}},\xi_x)$, where  $\xi_x$ is  the  component of the displacement vector illustrated.  \label{2107P2}}
\end{figure} 

\begin{figure}
	\hspace{-1.0cm}	
	\includegraphics[angle=0,width=1\columnwidth]{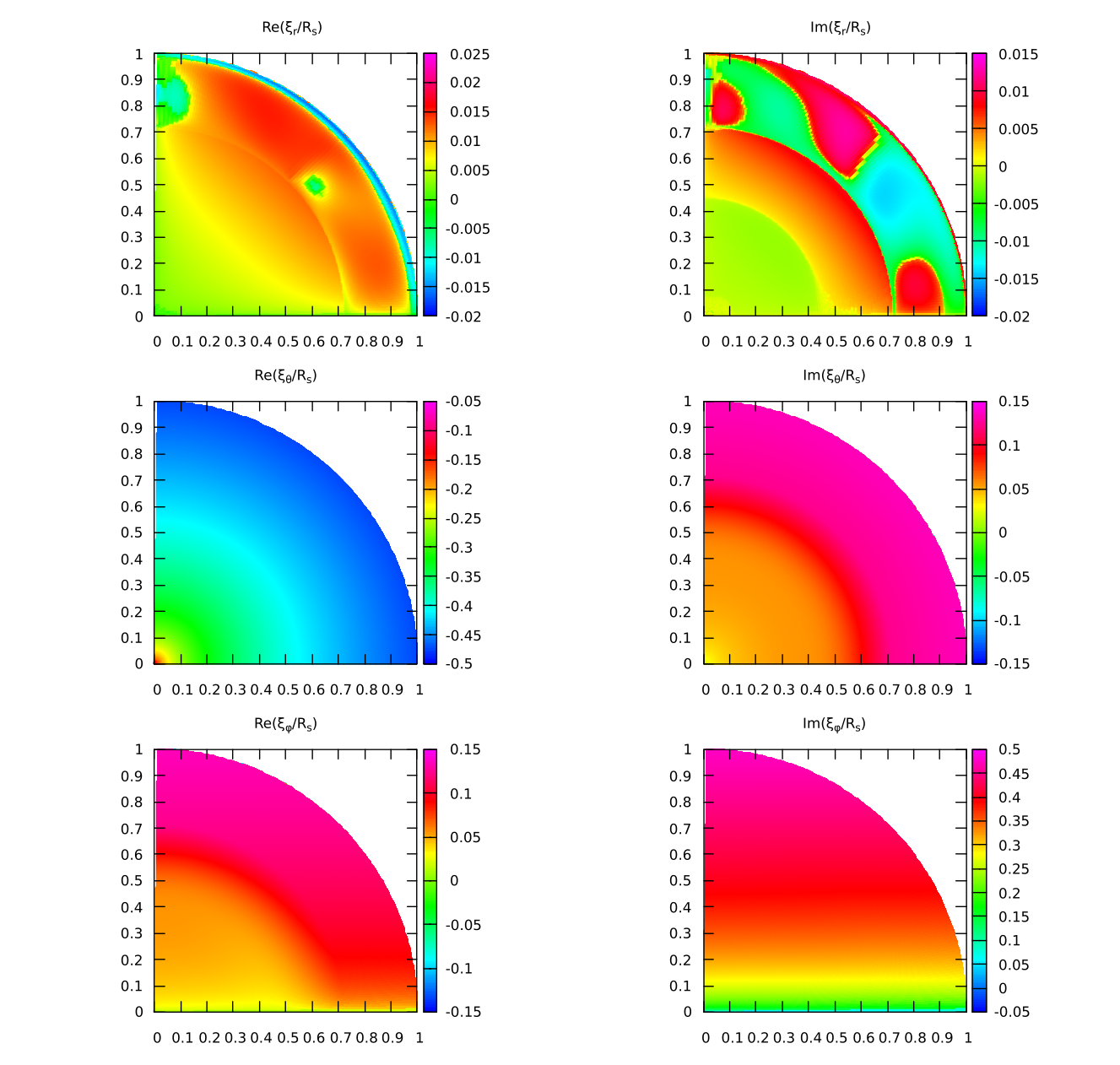}.
	\caption{{ As in Fig. \ref{2107P2} but showing contour plots  in the primary's meridional plane $\phi=0$ 
		illustrating the response to forcing with, ${\cal U}_{-1,0}, (n,m)= (-1,0)),$  for 
		 $\Omega_s=5\times 10^{-3}\Omega_c.$}
		 \label{2107P2a}}
\end{figure}

\begin{figure}
	\hspace{-1.0cm}	
	\includegraphics[angle=0,width=1\columnwidth]{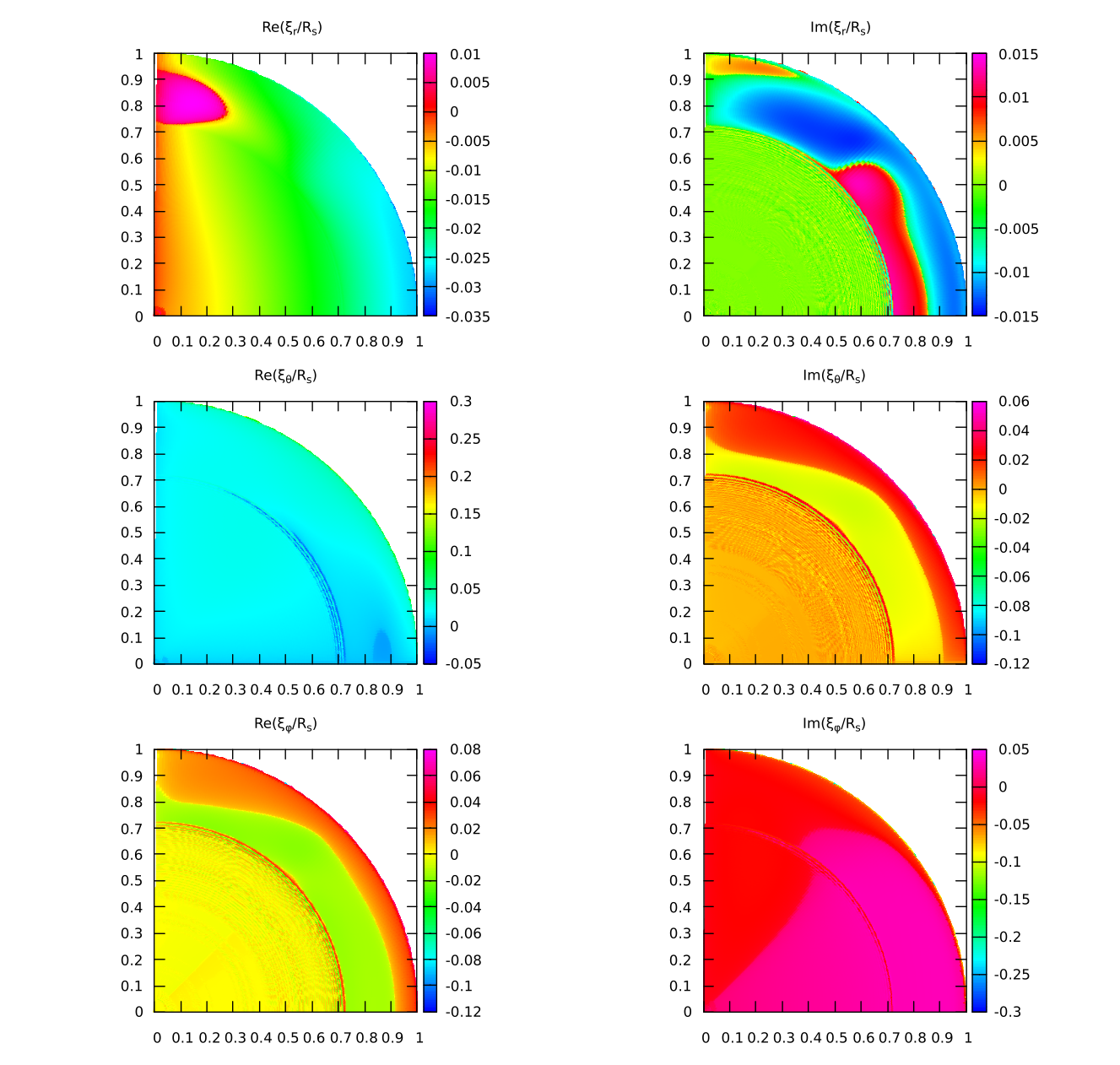}.
	\caption{{ Contour plots  in the primary's meridional plane $\phi=0$. These show the response to forcing with ${\cal U}_{-2,0},$ $(n,m)=(-2,0)$, for $\Omega_s=4.0\times10^{-3}\Omega_{c}$.
	The vertical colour bars on the right indicate the local value of  sign($|\xi_x|^{\frac{1}{4}},\xi_x)$, where  $\xi_x$ is  the  component of the displacement vector illustrated. } } \label{1207P5}
\end{figure} 

\begin{figure}
	\hspace{-1.0cm}	
	\includegraphics[angle=0,width=1\columnwidth]{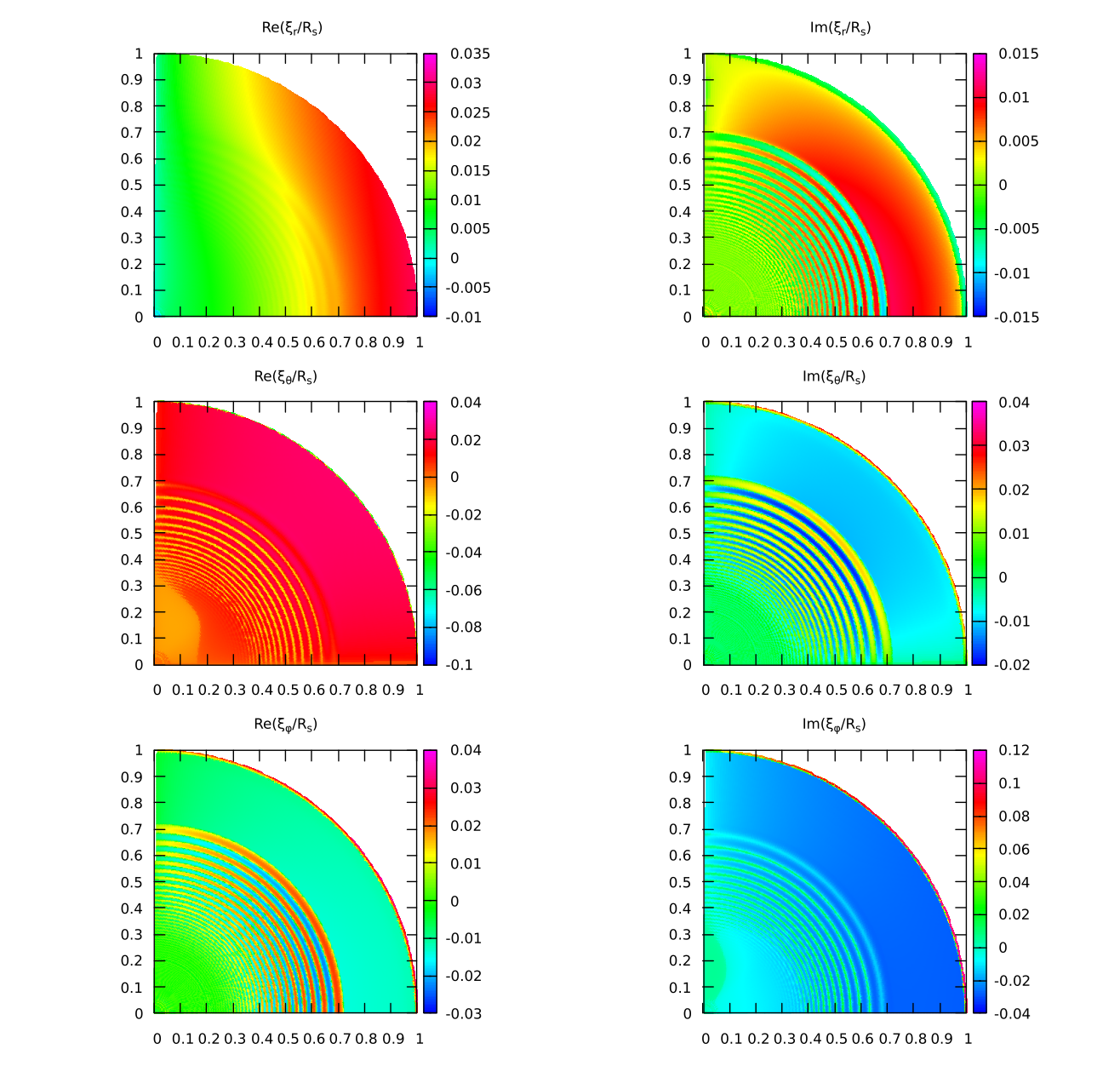}.
	\caption{{ As in Fig.\ref{1207P5} but showing contour plots  in the primary's meridional plane $\phi=0$ illustrating  the response to forcing with ${\cal U}_{-2,-2},$ $(n,m)=(-2,-2)$, for
	 $\Omega_s=4.0\times 10^{-3} \, \Omega_c$. 
	  In the radiative core  high radial order $g-$ modes are excited that are artificially damped in the inner core ($r/R_s < 0.3$) where the radial grid resolution is too poor.}
	 } \label{1207P5a}
\end{figure}

\section{Numerical calculation of the tidal response of the star and the viscous dissipation in its convective envelope}\label{Numerics}
We have performed tidal response calculations for a solar mass star with initial heavy element abundance $Z=0.02.$  The initial central hydrogen abundance  was $X_c=0.70.$
 The model was evolved  to  the stage when $X_c=0.40$,  when its radius had become $R_*=1.0051R_{\odot}$ and its luminosity $L_{*}=1.01L_{\odot},$ 
 these parameters being quite similar to those of  the current Sun. The calculations were  performed  with version r22.11.1 of the MESA stellar evolution code \citep{Paxton2015}. 

 We use this stellar model to calculate the tidal evolution of a solar type star and a hot Jupiter in close orbit.
 The convective envelope extends from $r/R_*= 0.727$ to $r/R_*=0.999$.
The moment of inertia of the star is $I =7.0413 \times 10^{53}$ g cm$^2$  and the critical stellar spin rate $\Omega_c = 6.2176 \times 10^{-4} s^{-1}$.
 When comparing the tidal response  of this model to a companion in a circular orbit  with $a =10 R_s$  to that obtained for a more evolved star (with $X_c=0.20$), we found very similar viscous dissipation rates while the stellar moment of inertia differed by  only a few percent. Hence we adopted the response  for the stellar model with $X_c=0.4$  to apply for further  main sequence stellar evolution.

 As in PS we consider a Jupiter mass companion in a circular orbit.  The numerical  procedures followed to obtain the tidal response in the star (generated by the orbiting planet) are described in detail 
 in PS and references therein. In solar type stars the tidal torque on the star is directly related to the viscous dissipation of the  kinetic energy of the tidal oscillations generated in the convective envelope. 
 Radiative dissipation in the radiative regions of the star is  expected to be much smaller and can be neglected (PS). 
 The  artificial viscous damping introduced in the radiative core to  deal with very short wavelength gravity waves, that would otherwise be unresolved (see below),   
  is  {not included in the computation of the} the stated viscous dissipation rates {on account of its resolution dependence.}

Following  PS a turbulent viscosity was assumed.  The kinematic viscosity
 $\nu(r)$  was taken from \cite{Duguid2020} as
\begin{equation}
	\nu(r)= 
	\frac{ {\frac{1}{3}}
	{ \mathcal{L}_{mx} v_{c} }}
	{(1+({\tau_{c}}/{P_{osc}})^{s})}
	\label{turbvisc}
\end{equation}
whereby the convective mixing length $ \mathcal{L}_{mx}= \alpha |H_P|$ is here scaled by the parameter $\alpha=2$. The local pressure scale height $|H_P(r)|$  and the local convective velocity $v_{c}(r)$ are taken from the  MESA input stellar  model. Any mismatch of the timescale of the forced oscillations ($P_{osc}=2 \pi/\omega_{f,n,m}$) and that of the turbulent convection ($\tau_c=1/\sqrt{|N^2|}$), where $N$ is the Brunt-V\"{a}is\"{a}l\"{a} frequency, is taken into account by the term in the denominator raised to the power $s=2.$

We introduced a thin layer with artificial viscosity in the transition layer between the radiative core and the inner boundary of the convective envelope at $r/R_*= r_{cb}/R_*=0.727$ by extrapolating the kinematic viscosity $\nu_{cb}$ at this boundary inwards according to \\
$\nu(r) = \nu_{cb} \exp \{-\left((r_{cb}-r)/(c_{*} H_P)\right)^2\} $
 up to the adopted minimum artificial viscosity of \\  $\nu_{min}=10^9$ cgs in the radiative core, where $H_P$ is here taken to be the pressure scale height at the convective boundary and the scaling factor is 
 $c_*=0.05$. In all cases, except for the forcing  with $n=-2$, the adopted minimum core viscosity of $\nu_{min}~=~10^9$ cgs gives sufficient damping to enable the determination and calculation of the $l'=1$ resonance for forcing  with $(n=-1,m=-2)$. However, forcing with $(n=-2, m=0)$  requires  $\nu_{min}=10^{10}$ cgs in the radiative core to damp grid oscillations of the displacement vector in the core.
 We comment that in the latter case $\nu_{min}/(R_*^2\Omega_c)\sim 3.25\times10^{-9}.$ Thus for the smallest angular velocity considered, namely, $\Omega_s=0.002\Omega_c,$ }
  the Ekman number $\nu_{min}/(R_*^2\Omega_s)\sim 1.6\times10^{-6}.$ As indicated above {the effect of this viscosity  is not included in the presented} total rate of dissipation which is  dominated by effects in the convection zone (see also  PS). 

As in PS the  viscous force  is derived from the viscous stress tensor for compressible flow $\bsf{\Sigma}_{i,j}$ expressed in spherical coordinates, whereby
 $\rho^{-1}\nabla \cdot \bsf{\Sigma}_{i,j} $
    gives the viscous force per unit mass 
    The viscous dissipation rate in the convective envelope follows by calculating\footnote{As this applies for a general forcing potential we drop the subscripts $n,m$ from $dE_{kin}/dt,$
    the stress tensor components, $\bsf{\Sigma}_{i,j},$ as well as the energy flux perturbations,  ${\bf F}'_{x},$ where $x \equiv c, rad,$ or $eq$  below  }
\begin{equation}
	\drv{E_{kin}}{t} = - \int_V \left( \frac{\frac{1}{2}\bsf{ \Sigma^*}_{r r} \bsf{\Sigma}_{r r} + \frac{1}{2} \bsf{\Sigma}^*_{\theta \theta} \bsf{\Sigma}_{\theta \theta} + \frac{1}{2} \bsf{\Sigma}^*_{\phi \phi} \bsf{\Sigma}_{\phi \phi} + \bsf{\Sigma}^*_{r \theta} \bsf{\Sigma}_{r \theta} +\bsf{\Sigma}^*_{\theta \phi} \bsf{\Sigma}_{\theta \phi} + \bsf{\Sigma}^*_{\phi r} \bsf{\Sigma}_{\phi r}} {\rho \nu} \right) dV \label{vdissip}
\end{equation}
where  the volume integral is taken over the convective envelope (up to $r=0.995R_*$ thus omitting the low mass superadiabatic surface layer).

 When applying the  frozen convection approximation, as was implemented in PS,  the tidal dissipation rate can become unrealistically large in the outer convective envelope
 where non adiabatic effects play a significant role in the evolution of the entropy variations induced by the tidal perturbation \citep[see][]{Bunting}.
 This effect was much  less significant for the calculations of PS which were mainly focused on $r$ mode resonances and their neighbourhoods, 
 with associated perturbations located mainly in the radiative core. 
 
  In the outer convective envelope the
 time scale associated with convective  energy becomes short enough 
 to result in the frozen convection  approximation becoming invalid. Unlike the situation when frozen, the convection is able to adjust and smooth out the rapid entropy
variations that would be produced by rapid variations of the divergence of the perturbed radiative flux and so reduce the amplitude of the response 
 \citep[see][]{Bunting}.  Taking this effect properly into account requires a rigorous treatment of convection that is currently unavailable.
 
 To proceed  we adopt a heuristic procedure based on physical arguments.
 We. assume  that  the   convective flux perturbation , ${\bf F}'_c,$  obeys a simple relaxation equation
\begin{equation}
	\partial{{\bf F}'_c}/\partial{t} = - \frac{( {\bf F}'_c-({\bf F}_{eq}'-{\bf F}'_{rad}))}{\tau_c},
\end{equation}
where  ${\bf F}'_{rad}$  and  ${\bf F}'_{eq}-{\bf F}'_{rad},$ are respectively the radiative  flux  and the  equilibrium convective  flux perturbations the relaxation process leads to,
with  $\tau_{c}$ being the convective time scale.  The quantity ${\bf F}'_{eq}$ is also the total flux the system would relax to in the limit of vanishing convective time scale.

For a forcing frequency, $\omega_{f,n,m},$
we thus have ${\bf F'}_{c} = ({\bf F}'_{eq} -{\bf  F}'_{rad})/(1+{\rm i}\omega_{f,n,m}\tau_{c}),$
and   
  \begin{equation}
  \nabla\cdot({\bf F}'_c +{\bf  F}'_{rad})=  \nabla\cdot(  ({\bf F}'_{eq} + {\rm i}\omega_{f,n,m}\tau_{c} {\bf  F}'_{rad})/(1+{\rm i}\omega_{f,n,m}\tau_{c})).\label{convrelax}
  \end{equation}
We assume that rapid variations in  ${\bf F}'_{rad}$ produced by density and temperature variations occur while the system relaxes towards a
slowly varying total flux, ${\bf F}_{eq}',$ such as would be expected for a near isentropic convection zone.  Accordingly we assume that its divergence may be neglected. 
In addition we also neglect variation of  $\tau_{c}.$ Accordingly (\ref{convrelax}) for the divergence of the total energy flux  perturbation  becomes
  \begin{equation}
  \nabla\cdot({\bf F}'_c +{\bf  F}'_{rad})= \frac{ {\rm i}\omega_{f,n,m}\tau_{c}\nabla\cdot {\bf  F}'_{rad}}{1+{\rm i}\omega_{f,n,m}\tau_{c}} .\label{convrelax1}
  \end{equation}

This procedure leads to the  application of  a complex reduction factor  $\epsilon_c$ in equation. (17) of PS  such that  $\nabla \cdot \boldsymbol{\mathcal {F}}'_{m,\sigma} \rightarrow  
\epsilon_c \nabla \cdot {\boldsymbol {\mathcal  F}'}_{m,\sigma}$ \footnote{This follows the notation of PS} where
$  \epsilon_c =	( \rm{i}\, \omega_{f,n,m} \,\tau_c + \omega_{f.n,m} ^2 \tau_c^2)/(1 + \omega_{f,n,m}^2 \tau_c^2).$
Notably this is unity in the limit of long convective time scale corresponding to frozen convection while producing, as expected from the above arguments,  a reduction factor for fluctuations
in the radiative flux when the convective time scale is short. 

\section{Numerical results}\label{Numres}
All numerical calculations of the stellar tidal response are calculated for a fixed semi-major axis $a = 10 \, R_*$, corresponding to an orbital period $P_{orb}=3.697 d$. The forcing frequency for $(n,m)$ forcing  in the rotating stellar frame is $\omega_{f,n,m}  = - m n_o ~+~ n \Omega_s $.  For $m=0$ forcing the tidal response depends on $\Omega_s$ only.  In this case the response for other values of $a$ can  be obtained by applying the scaling factor $(a/R_*)^3$ for the tidal perturbations and the factor $(a/R_*)^6$ for the viscous dissipation rate.
When $m\ne0$ and $n_o$ does not correspond to the above orbital period the same  scaling factors  may be applied to obtain results for the value of $a$ corresponding to the specified $n_o.$

\subsection{$r$ mode resonance}\label{rmoderes}
Toroidal or $r$ mode resonances are expected to  occur when   the forcing frequency  $\omega_{f,n,m}= -mn_o+n\Omega_s $  is close to $2n\Omega_s/l'(l'+1),$
where $l'$ is an integer (see PS) with the relative deviation $\rightarrow 0$ as $\Omega_s\rightarrow 0.$ This means that the precise location of a resonance can be  found for a small value of $n_o.$
The case of interest  here has $l'=1,$ with  $n=\pm 1$ together  with $m=\pm 2.$ Without loss of generality we focus on the case $n=-1,$ and $m=-2.$
The resonant frequencies are then  very close to $-\Omega_s$ which would be expected for a rigid tilt mode \textcolor{blue}{for which to within a constant of proportionality  
${\mbox{\boldmath$\xi$}}=(0,r, -{\rm i}r\cos\theta)\exp({\rm i}(\omega_{f,n,m} t- \phi)$}  (see PS).

The  dominant $l'=1$ $r$-mode resonance was calculated for a spin rate $\Omega_s / \Omega_c = 4 \times 10^{-3}$ by forcing with $n=-1$ and $m= -2$
 and zooming in on the resonance  by searching the forcing frequency for which the kinetic energy in the star becomes maximal (the procedure followed by PS). 
 The system's semi-major axis was  kept fixed at $a = 10 R_s$ noting the possibility of later scaling.
  
  The properties of the resonance are illustrated  in Fig.\ref{Disp-3107P2} which shows the resonance curves for the kinetic energy and the viscous dissipation.
As in PS these curves are fitted by the functional form  $ I_0/
(1+((\omega_f - \omega_0)/D)^2),$ Where the fitted parameters $I_0, \omega_0, $ and $D$ are indicated in Fig. \ref{Disp-3107P2}.
Contour plots for the tidal displacement
components  at the resonance frequency are  shown in Fig. \ref{c13.3107P2}. 
It will be seen that the displacement is mainly toroidal with $l'=1$ for which  $\xi_{\theta} $ is  independent of $\theta$ and $\xi_{\phi}\propto  \cos\theta. $
Thus for the horizontal components of the displacement  there are no nodes in $\theta.$
This {  corresponds to}  the dominant or fundamental mode with $n_r=0$ in the notation of PS. It is the mode with eigenfrequency in the rotating frame closest to $\Omega_s$ in magnitude.
However, it is not close to a rigid  tilt mode, which does not exist for our model, and
 for which $\xi_{\theta}$ and $\xi_{\phi}$ would be  $\propto r$ and thus have no nodes.
Fig. \ref{c13.3107P2}.  indicates at least one node \textcolor{blue}{ in both  $Re(\xi_{\theta})$ and $Im(\xi_{\phi}).$}

Note that the resonance width is much smaller than the magnitude of the deviation of  $|\omega_0|$
from $\Omega_s.$ This has the consequence that the calculations below, are  typically significantly into resonance wings.
{Here we remark that, although not included in the viscous dissipation calculation, the calculated resonance width is significantly increased by the presence of artificial viscosity
in the radiative core. As the mode is predominantly located in the radiative core, this causes the  profiles  illustrated in Fig. \ref {Disp-3107P2} to be significantly broadened. 
Although the structure of   the centre of the resonance, which in any case is likely to be affected by nonlinear effects (see PS) is modified, 
 the wings and consequently the discussion below are to a very good approximation unaffected.
 \footnote  {{Note similar broadening 
of resonance profiles occurred in PS on account of artificial viscosity in the outer radiative core employed to smooth the transition 
to the convective envelope.  As these were still very narrow, orbital evolution calculations are not significantly affected on account of 
rapid passage through  resonance centres. }}}

\subsection{ General tidal response for rotation rates such that $0.003\Omega_c  < \Omega_s < 0.010\Omega_c$}\label{Genresp}

As required for an application of equations (\ref{adotaveedot1}) and  (\ref{Gaussequation4.3z3}) to determine  the evolution of $a$ and $\beta,$ 
calculations   of the tidal response and consequent viscous dissipation rates  are presented  with forcing frequencies with $(n,m)=  (-1,0), (-2,0)$  and $(-2,-2).$
Discrete values of $\Omega_s/\Omega_c$ between $ 0.003$  and  $0.010$ 
corresponding to rotation periods between $39.0d$ and $11.7d$ were adopted, see tables \ref{tabl1}-\ref{tabl4}. 

\begin{table} 
	\begin{center}
		\begin{tabular} {ccccc} 
			\hline 
			$\Omega_s/\Omega_c$ & $ E_{kin,n,m}$ (erg) &   Dissipation rate (erg/s) & $\omega_{f,n,m}/\Omega_c$  & $P_s (d)$ \\ 
			\hline						
3.000000000E-03 & 5.130732E+40 & 1.278381E+24 &-3.000000E-03 & 3.898749E+01 \\
3.500000000E-03 & 3.756163E+40 & 9.188618E+23 &-3.500000E-03 & 3.341784E+01 \\
4.000000000E-03 & 2.865123E+40 & 6.888845E+23 &-4.000000E-03 & 2.924061E+01 \\
4.500000000E-03 & 2.255411E+40 & 5.352036E+23 &-4.500000E-03 & 2.599166E+01 \\
5.000000000E-03 & 1.820308E+40 & 4.276251E+23 &-5.000000E-03 & 2.339249E+01 \\
5.500000000E-03 & 1.498935E+40 & 3.475961E+23 &-5.500000E-03 & 2.126590E+01 \\
6.000000000E-03 & 1.254738E+40 & 2.855463E+23 &-6.000000E-03 & 1.949374E+01 \\ 
6.500000000E-03 & 1.064866E+40 & 2.368167E+23 &-6.500000E-03 & 1.799422E+01 \\
7.000000000E-03 & 9.143951E+39 & 1.983517E+23 &-7.000000E-03 & 1.670892E+01 \\
7.500000000E-03 & 7.932008E+39 & 1.677708E+23 &-7.500000E-03 & 1.559499E+01 \\
8.000000000E-03 & 6.942046E+39 & 1.432359E+23 &-8.000000E-03 & 1.462031E+01 \\
8.500000000E-03 & 6.123379E+39 & 1.233898E+23 &-8.500000E-03 & 1.376029E+01 \\ 
9.000000000E-03 & 5.438977E+39 & 1.071981E+23 &-9.000000E-03 & 1.299583E+01 \\
1.000000000E-02 & 4.369319E+39 & 8.258683E+22 &-1.000000E-02 & 1.169625E+01 \\
			\hline 			
		\end{tabular}
	\end{center}
	\caption{ Results for  forcing of the star with $(n,m)=(-1,0)$: the columns moving from left to right  contain the dimensionless 
	spin rate $\Omega_s/\Omega_c$, the kinetic energy,  the  Dissipation rate $-\dot{E}_{kin,n,m} \equiv - \dot{E}_{kin,-n-,m}  $, 
	the dimensionless forcing frequency $\omega_{f,n,m}/\Omega_c$ and spin period in days. }
	\label{tabl1}
\end{table}

\begin{table} 
	\begin{center}
		\begin{tabular} {ccccc} 
			\hline 
			$\Omega_s/\Omega_c$ & $ E_{kin,n,m}$ (erg) &   Dissipation rate (erg/s) & $\omega_{fn,m}/\Omega_c$  & $P_s (d)$ \\ 
			\hline
3.000000000E-03 & 9.670707E+30 & 5.922592E+23 &-6.000000E-03 & 3.898749E+01 \\
3.500000000E-03 & 7.518518E+30 & 4.859204E+23 &-7.000000E-03 & 3.341784E+01 \\
4.000000000E-03 & 6.264551E+30 & 4.195172E+23 &-8.000000E-03 & 2.924061E+01 \\
4.500000000E-03 & 5.591261E+30 & 3.776630E+23 &-9.000000E-03 & 2.599166E+01 \\
5.000000000E-03 & 5.360165E+30 & 3.516430E+23 &-1.000000E-02 & 2.339249E+01 \\ 
5.500000000E-03 & 5.398892E+30 & 3.347580E+23 &-1.100000E-02 & 2.126590E+01 \\
6.000000000E-03 & 5.564371E+30 & 3.233619E+23 &-1.200000E-02 & 1.949374E+01 \\
6.500000000E-03 & 5.777902E+30 & 3.146418E+23 &-1.300000E-02 & 1.799422E+01 \\
7.000000000E-03 & 5.803063E+30 & 3.052197E+23 &-1.400000E-02 & 1.670892E+01 \\
7.500000000E-03 & 5.934019E+30 & 3.001698E+23 &-1.500000E-02 & 1.559499E+01 \\
8.000000000E-03 & 6.359394E+30 & 2.995753E+23 &-1.600000E-02 & 1.462031E+01 \\
8.500000000E-03 & 6.977628E+30 & 3.008645E+23 &-1.700000E-02 & 1.376029E+01 \\
9.000000000E-03 & 7.585802E+30 & 3.037263E+23 &-1.800000E-02 & 1.299583E+01 \\
1.000000000E-02 & 9.008264E+30 & 3.148848E+23 &-2.000000E-02 & 1.169625E+01 \\
			\hline 			
		\end{tabular}
	\end{center}
	\caption{ Results for forcing of the star with $(n,m)=  (-2, 0)$ are tabulated :   the annotation is the same as for table  \ref{tabl1}}
	\label{tabl2}
\end{table}

\begin{table} 
	\begin{center}
		\begin{tabular} {ccccc} 
			\hline 
			$\Omega_s/\Omega_c$ & $ E_{kin,n,m}$ (erg) &   Dissipation rate (erg/s) & $\omega_{f,n,m}/\Omega_c$  & $P_s (d)$ \\ 
			\hline
3.000000000E-03 & 1.013433E+32 & 6.692503E+23 & 5.727574E-02 & 3.898749E+01 \\
3.500000000E-03 & 9.769323E+31 & 6.641119E+23 & 5.627574E-02 & 3.341784E+01 \\
4.000000000E-03 & 9.242478E+31 & 6.544582E+23 & 5.527574E-02 & 2.924061E+01 \\
4.500000000E-03 & 8.759001E+31 & 6.444667E+23 & 5.427574E-02 & 2.599166E+01 \\
5.000000000E-03 & 8.461381E+31 & 6.342059E+23 & 5.327574E-02 & 2.339249E+01 \\
5.500000000E-03 & 8.248848E+31 & 6.272568E+23 & 5.227574E-02 & 2.126590E+01 \\ 
6.000000000E-03 & 7.876073E+31 & 6.221043E+23 & 5.127574E-02 & 1.949374E+01 \\
6.500000000E-03 & 7.409705E+31 & 6.105802E+23 & 5.027574E-02 & 1.799422E+01 \\
7.000000000E-03 & 7.100741E+31 & 5.998191E+23 & 4.927574E-02 & 1.670892E+01 \\
7.500000000E-03 & 6.929987E+31 & 5.910451E+23 & 4.827574E-02 & 1.559499E+01 \\
8.000000000E-03 & 1.632116E+32 & 5.747212E+23 & 4.727574E-02 & 1.462031E+01 \\
8.500000000E-03 & 1.195258E+32 & 5.628202E+23 & 4.627574E-02 & 1.376029E+01 \\
9.000000000E-03 & 1.106304E+32 & 5.413451E+23 & 4.527574E-02 & 1.299583E+01 \\
1.000000000E-02 & 1.544881E+32 & 5.393494E+23 & 4.327574E-02 & 1.169625E+01 \\
			\hline 			
		\end{tabular}
	\end{center}
	\caption{Results for forcing of the star with $(n,m)=  (-2, -2)$ are tabulated:   the annotation is the same as for table  \ref{tabl1}}
	\label{tabl3}
\end{table}

For other values of $\Omega_s,$ \textcolor{blue}{and the corresponding $\omega_{f,n,m},$}   the viscous dissipation rates were obtained  by cubic spline \textcolor{blue}{interpolation/extrapolation},
  They are plotted  in Figs. \ref{Disprate1} and \ref{Disprate2}.
   These results  enable the Runge-Kutta time integration of the evolution equations of the system \citep{NumRec96} (see the next section). 
   
 {  The responses to  forcing with $(n,m)= (-1,0)$   for $\Omega_s= 3.0\times 10^{-3}\Omega_c.$
   and  $\Omega_s~=~5.0~\times~10^{-3}\Omega_c$  are illustrated by
   the contour plots in { Figs. \ref{2107P2} and \ref{2107P2a}} respectively.}
An $r$-mode response in the radiative core may potentially be excited. 
    Though, as explained above  the calculations presented here are significantly into the resonance wing.
   In spite of this {Figs. \ref{2107P2} and \ref{2107P2a}  indicate} that the perturbations still have a strong toroidal component.
   The viscous dissipation rate as a function of $\Omega_s/\Omega_c$  for $(n,m)= (-1,0)$  plotted in the { lower} panel of Fig. \ref{Disprate1} indicates
    that resonance is approached as $\Omega_s$ decreases.  
   
   Similar rates of dissipation are obtained for forcing with the component of the tidal perturbation with  $(n,m)=(-2,0).$ Contour plots  illustrating the response 
   for $\Omega_s= 4.0\times 10^{-3}\Omega_c.$ are provided in 
  {  Fig. \ref{1207P5}}.  In this case there is a strong  toroidal component with $l'=3$ as expected.

The forcing  frequencies $\omega_{f,-2,-2}$ for $(n,m)=(-2,-2)$  lie outside the inertial range and result in the excitation of $g$-modes of high radial order  in the radiative core
as indicated in the  contour plots in {  Fig. \ref{1207P5a}}. In this case the perturbation is mainly spheroidal with $l'=2$ for which  $\xi_{\phi} \propto \sin\theta.$
For $r/R_s < 0.3$ the response  could not be resolved and it was accordingly  artificially damped.

\begin{table} 
	\begin{center}
		\begin{tabular} {ccccc} 
			\hline 
			$\omega_{f,-2,-2}/\Omega_c$ & $ \Omega_s/\Omega_c$   & $P_s(d)$ & $\dot{S}_{-2,-2} $  &  $\dot{S}_{MB} $ \\ 
			\hline
 5.727574E-02 &  3.000000E-03 & 3.898749E+01 & 3.758618E+28 & -4.283183E+30 \\
5.627574E-02 &  3.500000E-03 & 3.341784E+01 & 3.796037E+28 & -6.801536E+30\\
5.527574E-02 &  4.000000E-03 & 2.924061E+01 & 3.808533E+28 & -1.015273E+31 \\
5.427574E-02 &  4.500000E-03 & 2.599166E+01 & 3.819487E+28 & -1.445574E+31 \\
5.327574E-02 &  5.000000E-03 & 2.339249E+01 & 3.829227E+28 & -1.982955E+31 \\
5.227574E-02 &  5.500000E-03 & 2.126590E+01 & 3.859718E+28 & -2.639313E+31 \\
5.127574E-02 &  6.000000E-03 & 1.949374E+01 & 3.902668E+28 & -3.426547E+31 \\ 
5.027574E-02 &  6.500000E-03 & 1.799422E+01 & 3.906561E+28 & -4.356553E+31 \\
4.927574E-02 &  7.000000E-03 & 1.670892E+01 & 3.915592E+28 & -5.441229E+31 \\
4.827574E-02 &  7.500000E-03 & 1.559499E+01 & 3.938239E+28 & -6.692474E+31 \\
4.727574E-02 &  8.000000E-03 & 1.462031E+01 & 3.910472E+28 & -8.122184E+31 \\
4.627574E-02 &  8.500000E-03 & 1.376029E+01 & 3.912250E+28 & -9.742259E+31 \\ 
4.527574E-02 &  9.000000E-03 & 1.299583E+01 & 3.846086E+28 & -1.156459E+32 \\
4.327574E-02 &  1.000000E-02 & 1.169625E+01 & 4.009000E+28 & -1.586364E+32 \\
			\hline 			
		\end{tabular}
	\end{center}
	\caption{ The tidal spin-up rate $\dot{S}_{-2,-2}= (-2/\omega_{f,-2,-2}) (\dot{E}_{kin,-2,-2})\equiv (2/\omega_{f,2,2})(\dot{E}_{kin,2,2})$ (assuming $\beta=0$) is found to be
	significantly smaller than the adopted  \citep{Skum72} magnetic spin down rate $\dot{S}_{MB}=-6.606 \times 10^{47} \, \Omega_s^3 $ for all spin rates considered.
	}
	\label{tabl4}
\end{table}

\subsection{Magnetic braking and tidal evolution of the system}\label{Magb}
The viscous dissipation rates  associated with  the obliquity tides with $(n,m)= (-1,0)$  and $(-2,0)$ tend to drive the  evolution towards alignment of $\bf{L}$ and $\bf{S}$. 
We remark that the energy dissipation rates provided in tables \ref{tabl1} -\ref{tabl3}  and Figs. \ref{Disprate1} and {\ref{Disprate2}
may be inserted into equation (\ref{Gaussequation4.3z30}) in order to obtain
the evolution rate of $\beta.$

Since $P_{orb}$ is  taken to be much smaller than $P_{spin}$ the  generated  tide with $(n,m)=(-2,-2)$  causes the stellar spin rate to increase. 
However, standard Skumanich magnetic braking \citep{Skum72} of the stellar spin is expected to  dominate that process
 (see table \ref{tabl4}).   Thus  this spin-down can in principle drive the system closer to an $r$-mode resonance with $l'=1$. 
  { The dominance of magnetic braking has the consequence that the tidal evolution is almost entirely driven by the responses with   $(n,m)= (-1,0)$  and $(-2,0)$.
 These are associated with  a forcing frequencies  $-\Omega_s$ and $-2\Omega_s$ respectively,  and thus the relationship to the  $r$ mode resonance  discussed above, 
 namely being located in its wings,  does not depend on the
 orbital period. In addition the affect  on the evolution of changing the latter,  to a very good approximation,  can be taken into account by scaling the amplitude
 of the tidal response. This is  in the same manner as changing the planet mass. In this way our calculations below undertaken for an initial orbital period of $3.697d$
 can be extended to apply to other values.
  A consequence  is that the rate of evolution changes.
  This aspect is discussed further in Section \ref{incMpef}.}
 
 Unfortunately, the physics of magnetic braking is complicated and not well understood. Magnetic braking in evolved solar type stars generally follows  the observationally derived  Skumanich expression \citep{Skum72} whereby the stellar spin angular momentum $S$ decreases as $\propto t^{-1/2}$ 
due to magnetic braking with $ {\dot S}=\dot{S}_{MB} = f_{MB} \, \Omega_s^3\equiv -\kappa S,$ defining the quantity $\kappa.$  The constant  $f_{MB}= -6.60 \times 10^{47}$ (cgs) is adapted to obtain the current solar spin period $P_{spin}=27$ d. Kepler observations of open clusters of known ages show solar type stars that follow  Skumanich  spin-down but there are also fast rotating stars that are not consistent with the standard magnetic braking expression \citep[see eg.][]{Gossage23}.

\begin{figure}
	\includegraphics[width=8cm]{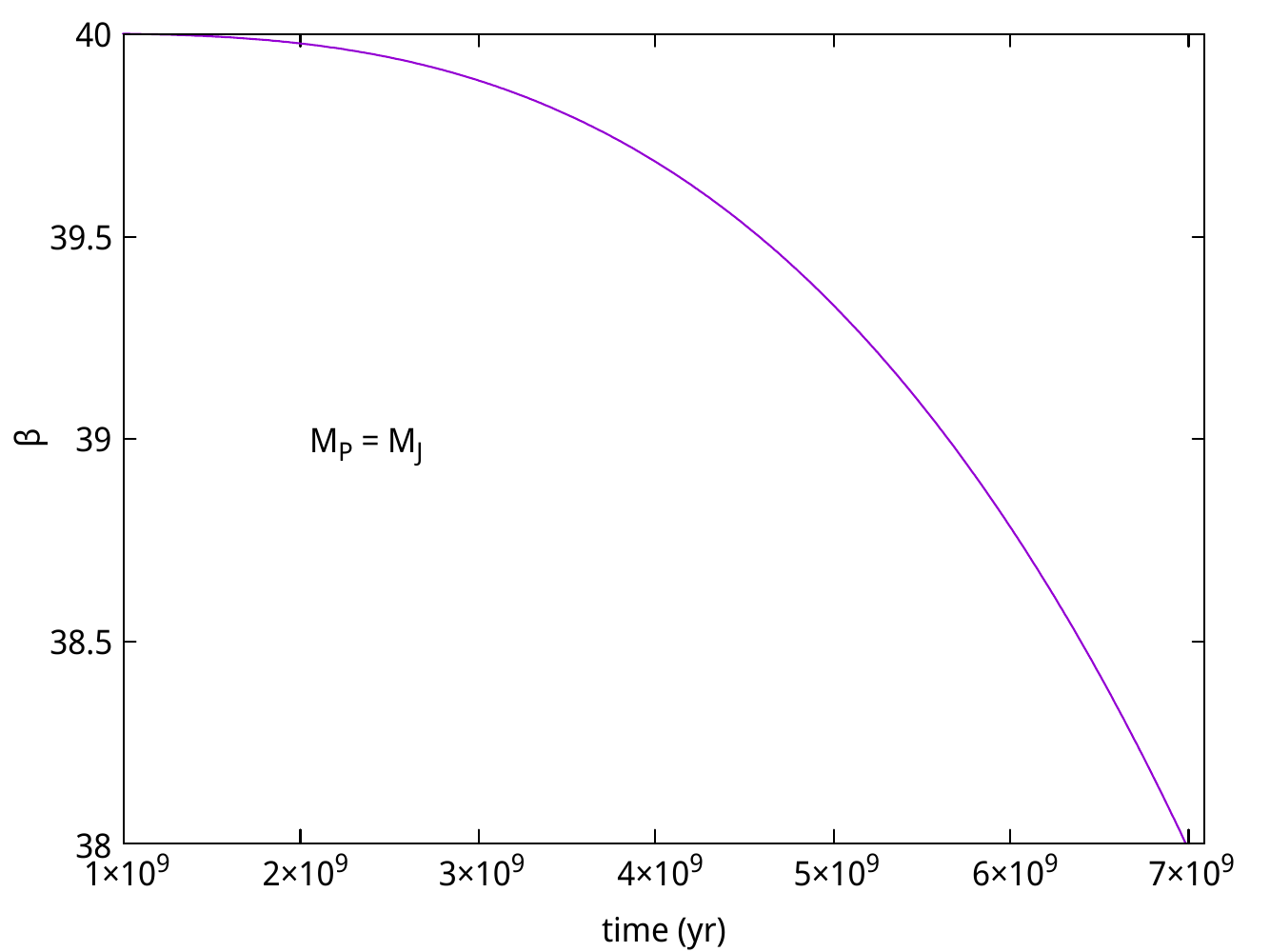}
	\includegraphics[width=8cm]{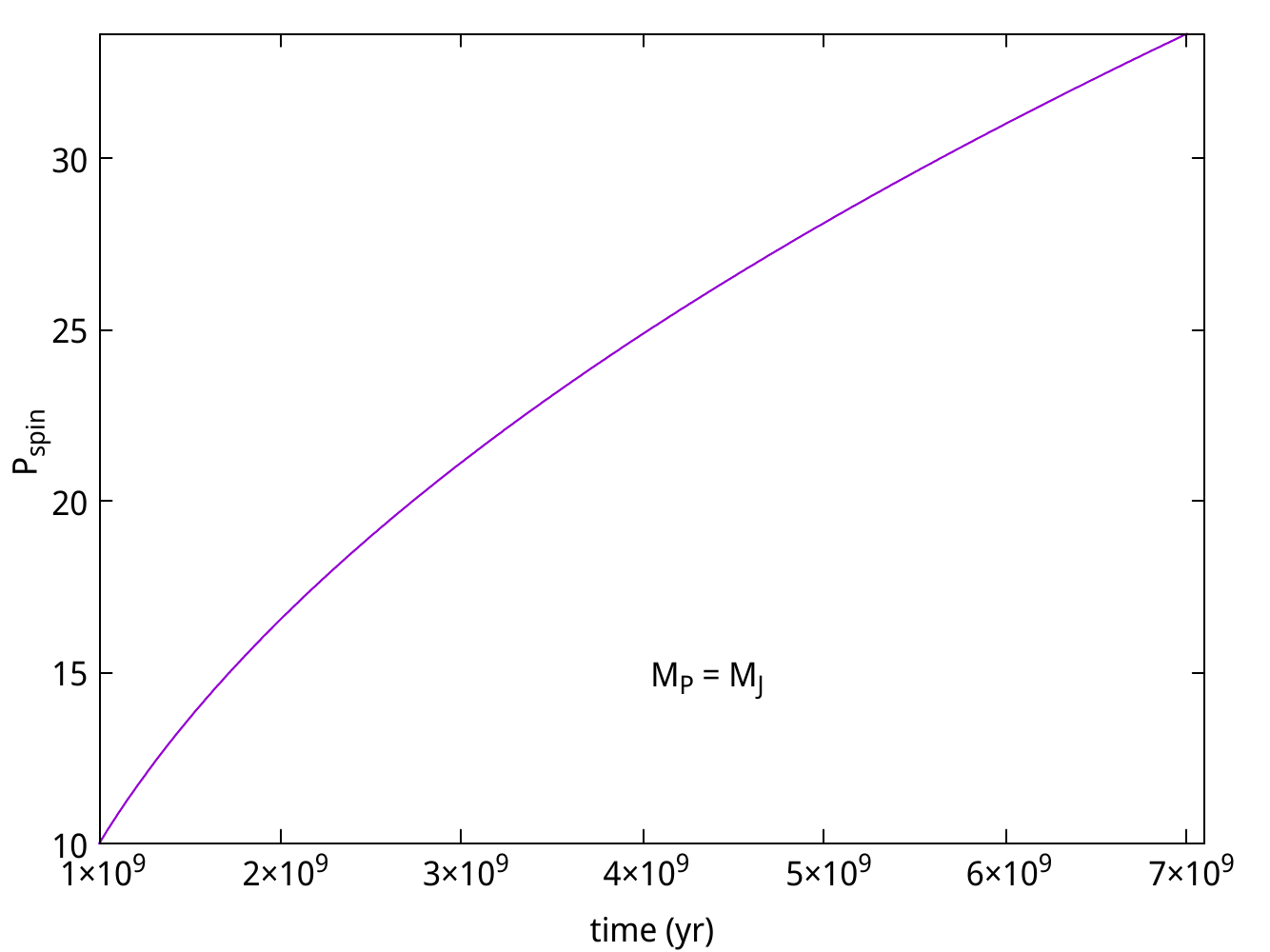}
	\caption{The left panel shows the time evolution of the angle $\beta$ (in degrees) between $\bf{S}$ and $\bf{L}$  driven by Skumanich magnetic braking with $f_{MB} = -6.60 \times 10^{47}$(cgs). The right panel shows the corresponding evolution of the stellar spin period in days. }  \label{MBcase1}
\end{figure}

\begin{figure}
	\includegraphics[width=8cm]{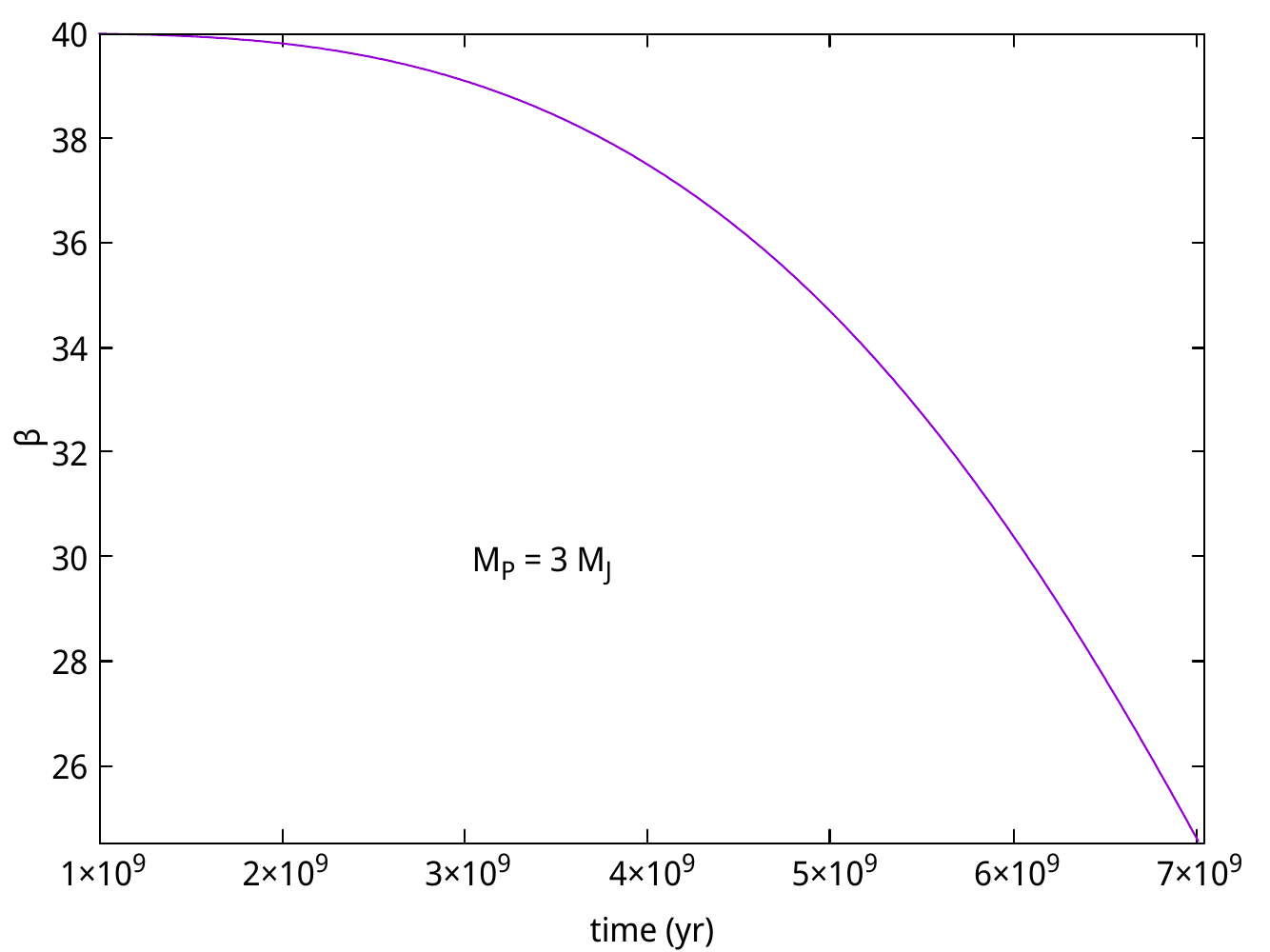}
	\includegraphics[width=8cm]{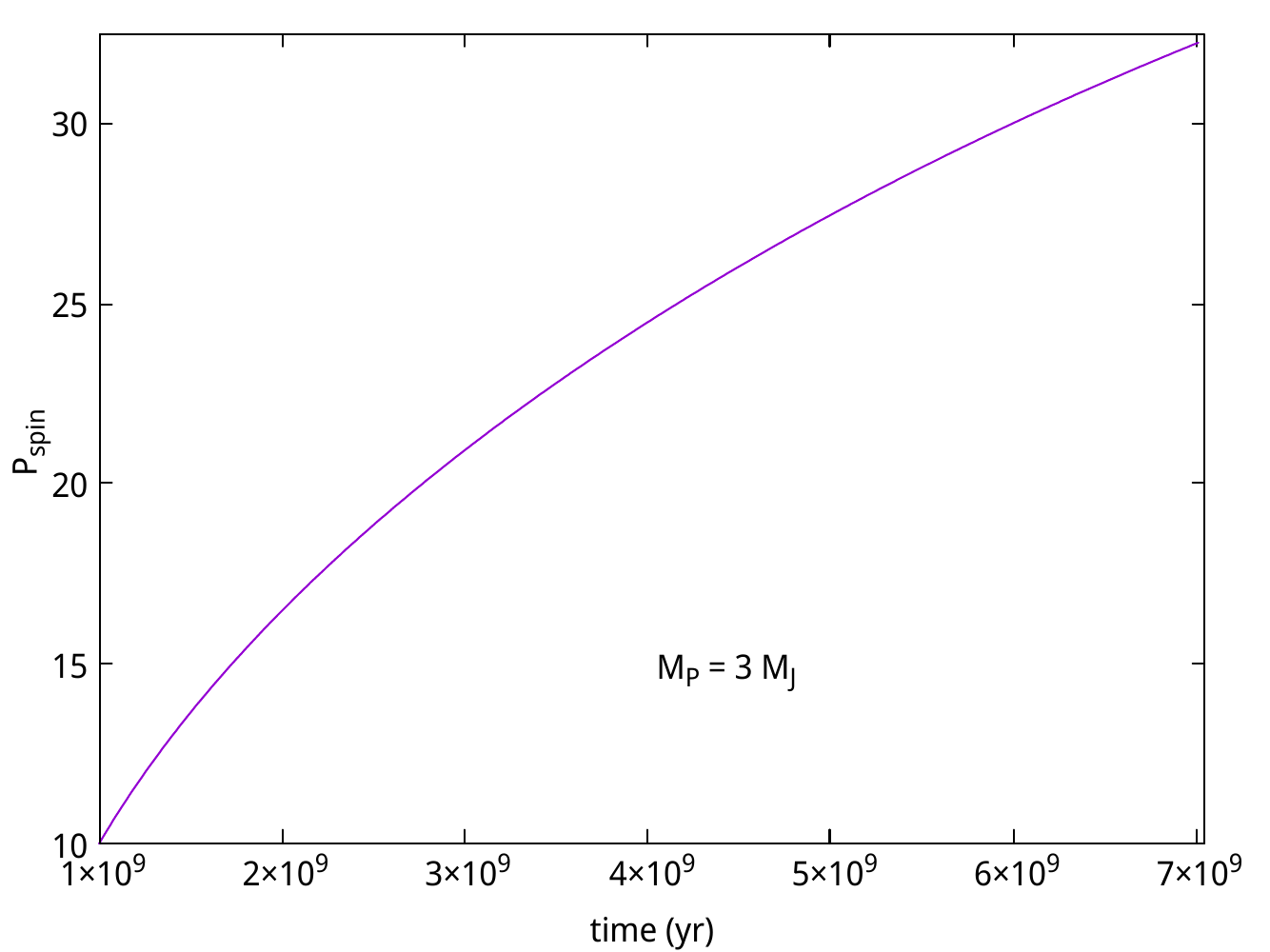}
	\caption{The left panel shows the time evolution of the angle $\beta$ (in degrees) between $\bf{S}$ and $\bf{L}$ driven by Skumanich magnetic braking  with $f_{MB} = -6.60 \times 10^{47}$ (cgs) for a planet of three Jupiter masses.  The right panel shows the corresponding evolution of the stellar spin period in days. }  \label{MBcase2}
\end{figure}

 \subsection{Numerical calculation of the evolution of the system}\label{Numcalcevol}
We follow the tidal evolution of the system by calculating the rate of change of $a$ given by equation (\ref{adotaveedot1})  and the rate of change of $\beta$ given by equation (\ref{Gaussequation4.3z30}).
We remark that  including magnetic braking is problematic as the angular momentum is no longer conserved as has been assumed above. 
Here we deal with this issue in a numerical treatment of the evolution by adopting an approach based on operator splitting \citep{Strang}.
Up to now, ${\bf L},$ and ${\bf S}$ obey equations of the form,
\begin{align}
 &d{\bf L}/dt= {\bf T}_{tidal}, \qquad d{\bf S}/dt= - {\bf T}_{tidal}, \hspace{3mm} {\rm and}  \hspace{3mm} d{\bf J}/dt= d({\bf L}+{\bf S})/dt= 0,\label{origset}
 \end{align}
where the last of these gives the conservation of angular momentum and  ${\bf T}_{tidal}$ is a tidal torque.
In order to incorporate spin down we modify the system to read 
\begin{align}
 &d{\bf L}/dt= {\bf T}_{tidal}, \qquad  d{\bf S}/dt= - {\bf T}_{tidal}-\kappa{\bf S} , \hspace{3mm} {\rm and}  \hspace{3mm} d{\bf J}/dt= d({\bf L}+{\bf S})/dt= -\kappa{\bf S}.\label{spinset}
\end {align}
Here we have introduced the spin down torque $-\kappa{\bf S}.$ 
To proceed by operator splitting we split (\ref{spinset}) into two systems. The first system is given by (\ref{origset})
and the second by
\begin{align}
 &d{\bf L}/dt= 0, \qquad  d{\bf S}/dt= -\kappa{\bf S} , \hspace{2mm} {\rm and}  \hspace{2mm} d{\bf J}/dt= d({\bf L}+{\bf S})/dt= -\kappa{\bf S},\label{spinset2}
\end {align}
The first system leads to equations 
(\ref{adotaveedot1})  and
(\ref{Gaussequation4.3z30})
with total angular momentum conservation as given above.
The second system is such that ${\bf S}$ preserves its direction while decreasing in magnitude. Thus $\beta$ does not change. Also ${\bf L}$ and hence $L$ do not change
Accordingly the right hand sides of 
(\ref{adotaveedot1})  and 
 (\ref{Gaussequation4.3z30})
 are identically zero and
we now have  
\begin {align}
\hspace{-2mm} \frac{{\bf J}}{J}\cdot \frac{d{\bf J}}{dt} =  -\frac{\kappa{\bf J}\cdot{\bf S}}{J} \hspace{2mm}  \rightarrow \hspace{2mm}  \frac{dJ}{dt}  =  -\kappa\frac{S}{J}(S+L\cos\beta)\label{spinset22}
\hspace{2mm} {\rm with} \hspace {2mm}J^2 = (S+L\cos\beta)^2+L^2\sin^2\beta.
\end {align}
We remark that  although it's magnitude changes ${\bf J}$ remains coplanar with ${\bf L}$ and ${\bf S}$ throughout .
\footnote{Using this result  we could start by taking, ${\bf J}_0,$ the initial value of ${\bf J}$  to define the $Z''$ axis of our coordinate system rather than ${\bf J}$ itself..
Note that ${\bf L}, {\bf S}. {\bf J}_0$ and ${\bf J}$ are all coplanar. The inclination $i$ will  then  be respect to ${\bf J_0}$ and this can
be shifted to be respect to ${\bf J}$ by taking into account the angle between ${\bf J}$ and ${\bf J}_0$. This approach yields identical conclusions as that of  the splitting approach. }

Now let the operator  $O_1(\Delta t)$ advance the first system (\ref{origset})  through a time step, $\Delta t,$ and the 
operator  $O_2(\Delta t)$ advance the second system (\ref{spinset2}) through the time step
$\Delta t.$ The splitting procedure advances a step, $2\Delta t,$ correct to second order, 
 by applying the sequence of operators, $O_1(\Delta t)O_2(2\Delta t)O_1(\Delta t)$ \citep{Strang}.
This procedure can be seen to be equivalent to solving the system of three equations 
(\ref{adotaveedot1})), (\ref{Gaussequation4.3z30}),  
and (\ref{spinset22}) directly. \textcolor{blue}{\footnote{.The same primary model is assumed throughout. Thus changes as a result of stellar evolution that were found to be small are neglected}           }
We implement it
using an adaptive step size controlled fifth order  Runge-Kutta \citep{NumRec96} subroutine RKQC to perform the first step  and subroutine RK4 to perform  the second step 
and advance $J$ using  (\ref{spinset22}).


In { Fig. \ref{MBcase1}} we  show the  time evolution of the angle $\beta$ for a solar type star and a planet with mass $M_P = M_J$ applying the Skumanich magnetic braking with $f_{MB}=-6.60\times 10^{47} cgs$
starting from an inclination $\beta=40^\circ.$
 It is found  that with this  induced  stellar spin-down rate 
 only a  small reduction of $\sim 2^\circ$  in $\beta$ occurs. 
 This rate of evolution is consistent with the discussion in Section \ref{testimate}. \textcolor{blue}{ As expected from the discussion in that Section}
   the orbital period \textcolor{blue} {is found to decrease  very slightly to}  3.683 d.
   The system  has not shifted sufficiently close to resonance with the $l'=1$ $r$-mode spectrum for larger changes to  $\beta$  to occur  during the Main Sequence phase.
        
    \subsection{Effect of increasing $M_p$ or decreasing $P_{orb}.$}\label{incMpef}
    In the context of these calculations we note that from 
  {  Fig. \ref{MBcase1}} 
    that for a rotation period, $P_{rot}= 33.6d,$ the angle $\beta$ decreases by $\sim 1^{\circ}$ in $10^9 y.$
    This is again consistent with the estimate in Section \ref{testimate}. 
    However, it is important to stress that these results apply to a planet with mass, $M_p,$  equal to one Jupiter mass with an orbital period of
    $3.697 d.$ From equation (\ref{Gaussequation4.3z30}) \textcolor{blue}{it follows that for fixed $M_*,$} the tidal evolution rate is $\propto M_p^2/P_{orb}^4.$
    Thus it will be increased by $\sim$ one order of magnitude if $M_p$ is increased by a factor of $\sim 3.$  Evolution of $\beta$  \textcolor{blue}{ would then become}  significant over the main sequence life time were  the rotation period to be maintained at $\sim 34d.$
   \textcolor{blue}{  Given that negligible change  to the orbital period is still  expected  during the tidal evolution, similar changes in $\beta$ would be expected were  the initial orbital period
    reduced by a factor  $\sim 1.75.$  }
    
    \textcolor{blue}{ In order to illustrate the above discussion  we  calculated the  
    the evolution of $\beta$ for a system for which the planet mass $M_P = 3 M_J$
    The results are illustrated in  { Fig. \ref{MBcase2}}.  In this case $\beta$  decreased from  $40^\circ$ to $24.5^\circ$ during the Main Sequence stage.  The orbital period decreased to 3.651 d.
   The corresponding evolution of the spin period (right panel in { Fig. \ref{MBcase2}}) shows that the stellar rotation period attained the value  $P_{spin} \simeq 32$ d.}

  \textcolor{blue}{In this context we remark that  \citet{Attia} note that the statistics of hot Jupiter missalingments indicate 
    increased significance of tidal effects for higher masses and shorter  orbital periods.}
     While stressing the uncertainties
     resulting from the use of a simplified stellar model as well as the crude treatment of convection,
    \textcolor{blue} {our results indicate that while it might have significant effects in some circumstances,
     the obliquity tide is unlikely to produce  strongly aligned hot Jupiter systems overall.}
\section{Discussion}\label{Discuss} 

In this paper we investigated the tidal interaction between a giant planet on a circular orbit around a solar type primary star.
Extending the work of PS, we considered the situation when the orbital and spin angular momenta were misaligned.
We obtained equations governing the inclination angle between the spin and orbital angular momenta $\beta$
and the semi-major axis which depended on the energy dissipation rates due to tidal perturbations
associated with forcing frequencies, $\omega_{f,n,m}= n\Omega_s - m n_o,$  with $(n,m)= (-1,0), (-2,0),$ and  $(-2,-2).$ 
We focused initially on the case where $M_p$ was one Jupiter mass and the orbital period was $3.697d.$

For the first of these \textcolor{blue}{ for which  $(n,m)= (-1,0),$} the perturbation  is stationary in the non rotating frame and  the spectrum of $r$ modes with $l'=1$
which has nearby eigenfrequencies may  potentially affect the response.
These modes have eigenfrequencies close to the frequency of a putative rigid tilt mode.  However, we recall that
such a mode does not exist for the model adopted apart from in the limit $\Omega_s=0.$
The properties of the fundamental $r$ mode ( with eigenfrequency closest to that of the putative tilt mode)
  were investigated in Section \ref{rmoderes}. We found that the resonance width was extremely small
such that even the small frequency mismatch associated with the forcing resulted in a response 
far into the wings. This was found to be the case for stellar rotation rates varying between  $11.8d$ and $39.0d$
(see table \ref{tabl1} 
 in conjunction with Fig.\ref{Disp-3107P2}).
 \textcolor{blue}{This can be understood in the following simple manner. Consider the case illustrated in Fig. \ref{Disp-3107P2}.
The dimensionless  resonant width is determined by the dissipation rate and, { even though it increased by artificial viscosity,} is $\sim  7.5\times 10^{-7}\omega_0$
( see Fig. \ref{Disp-3107P2} and PS). Whereas the relative frequency separation from the resonant frequency $-\Omega_s$ is 
$\sim (\Omega_s/\Omega_c)^2\sim 1.6\times 10^{-5} $ }
 This has the consequence that the full tidal response was essentially  non resonant
with the energy dissipation rates associated \textcolor{red} { with all relevant values of $\omega_{f,n,m}$ considered  being comparable.
}

Given this we  estimated  tidal evolution time scales from the governing equations in Section \ref{testimate} \textcolor{blue}{ obtaining
$|a (\langle da/dt \rangle)^{-1} | \approx 2.5\times 10^{12}y$ and
 $|\beta (\langle d\beta/dt\rangle)^{-1}| \sim 4\times 10^{10}y$ taking $M_p$  to be one Jupiter mass.
 The time scale for changing the orbital semi-major axis was  thus estimated as  a factor $\sim 60$ longer
than that for changing $\beta.$} 
These estimates were later confirmed by numerical. calculations of the orbital evolution
that in addition took into account magnetic braking in Section \ref {Numcalcevol}.

From equation (\ref{Gaussequation4.3z30}) the evolution of $\beta$ is towards $0$ for $\beta <\pi/2$
and towards $\pi$ for $\beta >\pi/2$ with an unstable stationary point at $\beta=\pi/2.$
Slow evolution in the neighbourhood of that point could result in a \textcolor{blue}{relative}  accumulation of systems
in near polar orbits. \textcolor{blue}{This could occur if tides are effective without an initial preference arising through the formation process.}  
 \citet{Albrecht} and \citet{Attia} have found statistical  evidence for such an accumulation.
However, \citet{Siegel} using a different approach do not find strong support for this at present.
This should be resolved by future work.

 The  spin up rate of the central star  induced by tides
was found to be very much less in magnitude  than the estimated spin down rate arising from magnetic braking
(see table \ref {tabl4}).  \textcolor{blue}{Thus} it is a reasonable approximation 
to consider the evolution of $\beta$ and $\Omega_s$. \textcolor{blue}{ to occur}  while the orbit remains fixed.

It can be seen from integrating  equation (\ref{Gaussequation4.3z30}) that  this  has the consequence \textcolor{blue}{ that if $\beta_0 < \pi/2 $  and $\beta_f$ are the initial and final
values of $\beta$ for the tidal evolution during} the  main sequence lifetime, for a given primary a    function ${\cal G}(\beta_0,\beta_f)$ which
 approaches the form $(\beta_f-\beta_0))/\beta_0$ as $\beta_f\rightarrow \beta_0$
 scales as $M_p^2/P_{orb}^4$.
 \textcolor{blue}{Notably in this approximation the scaling with $P_{orb}$ enables scaling to different orbital 
 periods without the need for further tidal response calculations 
 \textcolor{blue}{\footnote{ The case $\beta_0 > \pi/2$ can be considered through the mapping $\beta \rightarrow \pi-\beta$.}}.}

Thus  although we estimated from our calculations  that for a one Jupiter mass planet
and  a primary rotation period $\sim 33.6d,$ $\beta$  would change by about $1^{\circ}$ in $10^9 y,$
this would increase by about an order of magnitude for $M_p\sim 3$ Jupiter masses.
Thus the decrease of $\beta$ of from $40^{\circ}$ by about $2^{\circ}$ we found for a one  Jupiter mass planet 
over a main sequence lifetime as illustrated in { Fig.\ref{MBcase1}}  increases to $\sim 15^{\circ}$
for $3$ Jupiter masses. This is consistent with the finding of evidence  supporting the increased efficacy of tides
for larger planetary masses and shorter orbital periods by \citet{Attia}.

\textcolor{blue}{The  alignment distribution of close in giant planets is potentially  determined by a multitude of processes
affecting  individual objects in different ways \citep[see eg.][]{Siegel, Wrightetal, Wuetal}.
Some planets may undergo quiescent disc migration with only  modest dynamical interactions leading to modest spin-orbit  misalignment.
Others may  undergo more violent interactions leading to larger misalignments.
It has been suggested that tidal interactions may be responsible for greater  alignment of planets around cool stars  but not hotter stars beyond the Kraft break
on account of the lack of an envelope convection zone \citep[eg.][]{Albrecht12}. However, strong dynamical interactions may be preferred for 
stars beyond the Kraft break \citep{ Wrightetal, Wuetal}. In addition the distribution of warm Jupiter misalignments 
for which tidal effects are expected to be ineffective indicates a quiescent formation process can occur.}

\textcolor{blue} {Although not obviously required ab initio,} the discussion of \citet{Attia} indicates some  influence of tides, though this seems to be modest.
In support of that view the results presented here tend to indicate a potentially significant influence of tides, but only for giant
planets with  relatively large masses and  short orbital periods.


However, the limitations and uncertainties associated with our results need to be emphasised.
These relate to the use of a simplified stellar model that neglected centrifugal distortion,
being equivalent to one immersed in a fixed background potential designed to cancel out the
centrifugal potential. The  tidal forcing with $(n,m)=(-1,0)$ may be more strongly affected by  resonance with
$r$ modes leading to faster tidal evolution in a more realistic model.  In addition there are
significant uncertainties associated with the effective viscosity arising from  convection.
These are issues to be addressed in future work.

\section{Data Availability}
The data underlying this article will be shared on reasonable request to the corresponding author.

\begin{appendix}

\section{ Elements of the Wigner ${\bf {\lowercase { d}} }$ matrix}\label{Wigner}
These are standard and the elements of interest in our case are given by 
(see e.g. \cite{KMV})
\begin{align}
&d_{2,2}= \frac{1}{4}(1+\cos\beta)^2,\nonumber\\
&d_{2,1}= -\frac{1}{2}\sin\beta(1+\cos\beta),\nonumber\\
&d_{2,0}= \sqrt{\frac{3}{8}}\sin^2\beta,\nonumber\\
&d_{2,-1}= -\frac{1}{2}\sin\beta(1-\cos\beta),\nonumber\\
&d_{2,-2}= \frac{1}{4}(1-\cos\beta)^2,\nonumber\\
&d_{1,1}= \frac{1}{2}(2\cos^2\beta+\cos\beta -1),\nonumber\\
&d_{1,0}= -2\sqrt{\frac{3}{8}}\sin\beta\cos\beta,\nonumber\\
&d_{1,-1}= \frac{1}{2}(-2\cos^2\beta+\cos\beta +1),\nonumber\\
&d_{0,0}= \frac{1}{2}(3\cos^2\beta -1).
\end{align}
Note that for ease of notation we have dropped the superscript $2$ and that components not listed can be obtained from those listed by making use of 
the relations $d_{n_1,n_2}(\beta) =(-1)^{n_1-n_2} d_{-n_1,-n_2}(\beta)$
and $d_{n_1,n_2}(\beta) = d_{n_2,n_1}(-\beta).$

\end{appendix}
\end{document}